\documentclass[10pt, journal, letterpaper]{IEEEtran}
\usepackage{subfigure}
\usepackage{graphicx}
\usepackage{epsfig}
\usepackage{psfrag}
\usepackage{pstricks}
\usepackage{amssymb}
\usepackage{color}
\usepackage{algorithm}
\usepackage[noend]{algorithmic}
\usepackage{url}
\usepackage{graphicx, color}
\usepackage{balance}
\usepackage{cite}
\usepackage{amsmath}

\newcommand{\paren}[1]{\left(#1\right)}

\newcommand{\brparen}[1]{\left\{#1\right\}}
\newcommand{\abs}[1]{\left| #1\right|}
\newcommand{\field}[1]{\ensuremath{\mathbb{#1}}}
\newcommand{\R}{\ensuremath{\field{R}}} 
\newcommand{\I}[1]{\ensuremath{\mathsf{1}_{\left\{#1\right\}}}} 
\newcommand{\PR}[1]{\ensuremath{\mathsf{Pr}\left\{#1\right\}}} 
\newcommand{\EW}{\ensuremath{\mathsf{E}}} 
\newcommand{\ES}[1]{\ensuremath{\mathsf{E}\left[#1 \right]}} 
\newcommand{\ESI}[2]{\ensuremath{\mathsf{E}_{#2}\left[#1 \right]}}
\newcommand{\defeq}{\ensuremath{\triangleq}} 
\newcommand{\vecbold}[1]{\ensuremath{\boldsymbol{#1}}}

\newcommand{\on}{\ensuremath{{\rm ON}}}
\newcommand{\off}{\ensuremath{{\rm OFF}}}
\newcommand{\hist}{\ensuremath{{\mathcal{H}}}}
\newcommand{\cost}[2]{\ensuremath{J_{\vec{\pi}}^{\paren{#1, #2}}\paren{\vec{x}_0, \tau_0}}}
\newcommand{\mincost}[2]{\ensuremath{J^{\star\paren{#1, #2}}}\paren{\vec{x}_0, \tau_0}}

\newcommand{\gocost}[3]{\ensuremath{J_{#3}^{\paren{#1, #2}}\paren{\vec{x}_{#3}, \tau_{#3}}}}
\newcommand{\gocostON}[3]{\ensuremath{J_{#3}^{\paren{#1, #2}}\paren{\vec{x}_{#3}, 1}}}
\newcommand{\gocostOFF}[3]{\ensuremath{J_{#3}^{\paren{#1, #2}}\paren{\vec{x}_{#3}, 0}}}
\newcommand{\mingocost}[3]{\ensuremath{J_{#3}^{\star \paren{#1, #2}}\paren{\vec{x}_{#3}, \tau_{#3}}}}
\newcommand{\mingocostON}[3]{\ensuremath{J_{#3}^{\star \paren{#1, #2}}\paren{\vec{x}_{#3}, 1}}}
\newcommand{\mingocostOFF}[3]{\ensuremath{J_{#3}^{\star \paren{#1, #2}}\paren{\vec{x}_{#3}, 0}}}

\renewcommand{\vec}[1]{\ensuremath{\boldsymbol{#1}}} 

\newcommand{\tr}[1]{\ensuremath{{\rm Tr}\left(#1\right)}}

\usepackage[utf8]{inputenc}
\usepackage{amsmath}

\newtheorem{theorem}{Theorem}

\newtheorem{definition}{Definition}

\IEEEoverridecommandlockouts
\begin{document}

\title{Virtualized Control over Fog: Interplay Between Reliability and Latency}

\author{Submission \# 1570386822 \vspace*{-0.0cm}}
\author{Hazer Inaltekin, \emph{Member, IEEE}, Maria Gorlatova, \emph{Member, IEEE}, Mung Chiang, \emph{Fellow, IEEE}.\thanks{H. Inaltekin is with the Department of Electrical and Electronic Engineering, University of Melbourne, Parkville, VIC 3010, Australia (e-mail: hazeri@unimelb.edu.au).

M. Gorlatova is with the Department of Electrical Engineering, Princeton University, Princeton, NJ 08544, USA (e-mail: mariaag@princeton.edu).

M. Chiang is with the School of Electrical and Computer Engineering, Purdue University, West Lafayette, IN 47907, USA (e-mail: chiang@purdue.edu).}}
\maketitle

\vspace*{-0.0cm}
\begin{abstract}
This paper introduces an analytical framework to investigate optimal design choices for the placement of virtual controllers along the cloud-to-things continuum. The main application scenarios include low-latency cyber-physical systems in which real-time control actions are required in response to the changes in states of an IoT node. In such cases, deploying controller software on a cloud server is often not tolerable due to delay from the  network edge to the cloud. Hence, it is desirable to trade reliability with latency by moving controller logic closer to the network edge. Modeling the IoT node as a dynamical system that evolves linearly in time with quadratic penalty for state deviations, recursive expressions for the optimum control policy and the resulting minimum cost value are obtained by taking virtual fog controller reliability and response time latency into account. Our results indicate that latency is more critical than reliability in provisoning virtualized control services over fog endpoints, as it determines the swiftness of the fog control system as well as the timeliness of state measurements. Based on a realistic drone trajectory tracking model, an extensive simulation study is also performed to illustrate the influence of reliability and latency on the control of autonomous vehicles over fog.
\end{abstract}

\vspace*{-0.0cm}


\section{Introduction} 

Fog computing, sometimes referred to as edge computing, is an emerging computing paradigm in which computing, storage, networking and control are placed at multiple locations between the endpoint devices and the cloud~\cite{bonomi2012fog,Chiang2016Fog}. Peter Levine, a partner at an A-list venture capital firm Andreessen Horowitz, has recently called fog computing \emph{the next multi-billion dollar tech market}~\cite{Levine2016End}. The promise of fog computing for enabling the next generation of advances in IoT is underscored by the growing developments of fog computing architectures~\cite{AmazonGreenGrass,AzureEdge2017} and ongoing industry-wide standardization efforts~\cite{OpenFogReferenceArchitecture2017,ETSI_MEC2017}.

Fog computing offers 
flexibility in the choice of {\em virtualized} controller placement options for interactive control applications, as has been proposed in the outlines of the vision of the future of the industry
\cite{bonomi2012fog, Chiang2016Fog, OpenFogReferenceArchitecture2017}. Furthermore, fog/cloud architecture is also starting to be considered from a practical point of view for futuristic control applications, e.g., moving vehicular controls to different locations is proposed in \cite{Esen15, Qi2016Design}. However, while multiple virtual controller placements are starting to become possible in practice \cite{Abdelaal17, Vick16, Yannuzzi2017Cities, Faruque16}, the theoretical foundations for these placement decisions are currently lacking. We take steps towards addressing this gap in the current paper. 

\begin{figure}[t]
  \centering
  \includegraphics[width=0.85\linewidth]{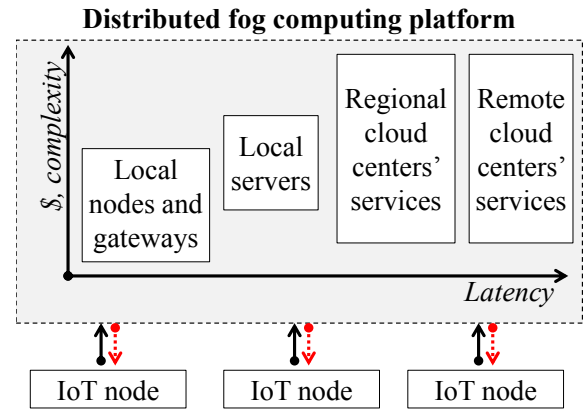}
 \vspace{-0.2cm}
  \caption{In a fog computing system, control application can be run at different distributed points and as different services, with different characteristics.\label{fig:FogFabic}}
  \vspace{-0.0cm}
\end{figure}

In particular, this paper focuses on \emph{latency} and {\em reliability} aspects that arise in a fog computing environment because different virtual controller locations in a fog hierarchy may exhibit different latency and reliability characteristics \cite{Chiang2016Fog}. For example, fog logic execution points may include local nodes and a wide variety of remote ones, as shown in Fig.~\ref{fig:FogFabic} (i.e., both Amazon Web Service (AWS) Greengrass \cite{AmazonGreenGrass}
and Microsoft Azure IoT Edge \cite{AzureEdge2017}
allow executing functions both locally and remotely). In these settings, local devices provide low response latency but may not always be reliable. Remote cloud computing nodes, on the other hand, offer considerably longer response times \cite{Gao2015Cloudlets} but can be readily designed to guarantee high reliability.
Then, what is the optimum design choice for placing controller software to maximize system performance? Critical to resolving this question is the discovery of the interplay between latency and reliability in control applications over fog, which is what the current paper achieves for linear IoT systems with a  quadratic cost.

\begin{figure}[!t]
  \centering
  \includegraphics[width=0.75\linewidth]{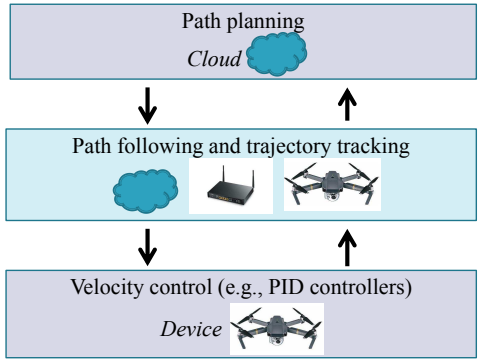}
    \vspace{-0.2cm}
  \caption{Layers of control functionality for drones. While path planning belongs on the cloud and velocity control probably belongs on the device itself, trajectory tracking elements could be virtualized on different fog endpoints. \label{fig:UAVControl}}
  \vspace{-0.2cm}
\end{figure}

Our analytical framework applies to all IoT systems with linear feedback controllers, which are studied in a wide
variety of applications \cite{astrom2010feedback, Bertsekas95}, that can be virtualized over the fog endpoints.  In particular, the trajectory following control for flying drones is a notable example of a control functionality that can be virtualized in different locations in a fog computing system, as shown in Fig.~\ref{fig:UAVControl}, and hence our results can be applied to. Motivated by the advances in quadcopter technology and by the commercial promise of autonomous drone operations, such as Amazon's plan to deliver packages using drones \cite{AmazonAir2017}, various aspects of drone operations are actively studied \cite{Khan2017Information,Wang2016Skyeyes}.
In drone air traffic control, while the highest-level global fleet planning decisions require the involvement of the cloud and low-level high-bandwidth velocity control needs to be done on the drone itself \cite{Bregu2016Reactive}, the important trajectory tracking and path following control operations~\cite{Sujit2013PathFollowing,Kukreti2016Genetically} can be executed on multiple locations in a fog network.


To the best of our knowledge, this paper is the first systematic study to shed light on the interplay between reliability and latency appearing in virtual control services offered over fog networks. Our main contributions can be summarized as follows.
\begin{itemize}
\item We propose an analytical framework to investigate the effects of latency and reliability on controlling linear IoT processes, disturbed by stochastic environmental factors, by means of a control software located along the cloud-to-things continuum. Under this framework, the min-cost performance of virtualized control services is obtained.
\item We derive the structure of optimum virtual controllers by considering reliability and latency (both communication and computation) characteristics of the fog endpoint which will execute the controller application. In addition to increased response times between consecutive control actions, we show that an estimator, separated from control, must first be run for distant fog endpoints to estimate live IoT node states from delayed sensor inputs. This collateral effect of latency further decreases the efficacy of software-defined control over imperfectly placed fog endpoints.
\item Based on a realistic drone trajectory tracking model, we conduct extensive simulations to visualize the performance of virtual fog controllers. It is observed that the path following efficiency decreases more quickly with latency than reliability due to its direct and collateral effects (i.e., increased response times and state estimation problem), which suggests moving controller software as close as possible to the unmanned vehicle (UV).
\end{itemize}

The remainder of this paper is organized as follows. In Section \ref{Section: Related Work}, we compare and contrast our results with related work.  In Section \ref{Section: Architectures}, we elaborate on important properties of fog computing architectures, and present small-scale results related to latency and reliability in fog.  In Section \ref{Section: System Model}, we introduce the analytical framework to investigate reliability and latency for virtualized control services over fog. In Section \ref{Section: Control Without Latency}, we derive the structure of optimum virtual controller without latency, while Section \ref{Section: Control With Latency} contains parallel results for optimum virtual controller with latency. In Section \ref{Section: Numerical Results}, we present our numerical results for the UV trajectory tracking problem. Section \ref{Section: Conclusions} concludes the paper, with potential generalizations of our results. 





\section{Related Work} \label{Section: Related Work}


Our results in this paper are related both to the emerging body of papers in fog computing \cite{Yannuzzi2017Cities, Faruque16, tong2016hierarchical, Tan2017Online, Xiao2017QoE, Kosta12} and to the more classical literature in control systems \cite{Abdelaal17, Vick16, Liberatore2006Integrated, Ploplys2004Closed, Gatsis2015Control, Pappas15, Yuksel06, Gupta07}. The papers \cite{Yannuzzi2017Cities, Faruque16} focused on the development of fog computing platforms for smart-city and smart-home applications with several control functionalities virtualized either in street cabinets \cite{Yannuzzi2017Cities} or at the control panel located inside a home \cite{Faruque16}. Although these papers provide insightful system implementation showcases to illustrate the utility of fog computing, they do not take any analytical approach, as we do in this paper, to substantiate their design choices.

The papers \cite{tong2016hierarchical, Tan2017Online, Xiao2017QoE, Kosta12} studied how to adapt services for fog computing by mainly focusing on computational offloading and associated computing job scheduling. In particular, Tong et al. proposed a hierachical edge/cloud architecture in \cite{tong2016hierarchical}, and showed that the proposed architecture has a higher chance of serving peak loads from virtualized services. In \cite{Tan2017Online}, the authors studied online algorithms for minimizing total weighted response time for edge-cloud networks with upload and download delays. An important feature of their online algorithm is that its performance comes close to the optimal offline algorithm with speed augmentation and without requiring any ex-ante knowledge of job arrival statistics. In \cite{Xiao2017QoE}, the authors investigated a job offloading problem similar to those studied in \cite{tong2016hierarchical, Tan2017Online}, but by considering interest of both fog endpoints and users. Specifically, they optimized reponse times subject to power efficiency constraints of fog nodes and showed that cooperation among fog nodes has the potential to improve service execution times. In \cite{Kosta12}, Kosta et al. developed a novel mobile cloud computing platform to migrate smartphone applications to virtual machines running on the cloud in an attempt to improve mobile computing and energy efficiency at the network edge.

When compared to \cite{tong2016hierarchical, Tan2017Online, Xiao2017QoE, Kosta12}, we take a more fundamental approach in this paper. By focusing on virtual control services, we examine the problem of where to place the codebase for a single controller application along the cloud-to-things continuum. For each design choice of reliability and latency, we obtain the min-cost performance of the optimum virtual fog controller. These optimum performance values can then be inputed to a wider system-level fog optimization problem as in \cite{tong2016hierarchical, Tan2017Online, Xiao2017QoE} to determine how to dispatch and schedule a multitude of virtual control services to maximize a collective system utility, which we plan to pursue as a future research direction.

On the side of control systems, the papers \cite{Abdelaal17, Vick16} considered virtualized control services over the cloud. In \cite{Liberatore2006Integrated}, the author investigated an integrated play-back mechanism to improve the efficiency of remote control over the network. In \cite{Ploplys2004Closed}, they studied the design of a physical control system over a wireless link that can corrupt transmitted data. These papers, however, do not employ any optimization framework to compute the structure of a virtual controller to be run at a fog endpoint. The papers \cite{Gatsis2015Control, Pappas15, Yuksel06, Gupta07}, on the other hand, adopted a more optimization theory based approach to design control systems over communication channels either allowing opportunistic transmissions \cite{Gatsis2015Control, Pappas15} or dropping packets randomly \cite{Yuksel06, Gupta07}. The main point of difference of the current paper from \cite{Gatsis2015Control, Pappas15, Yuksel06, Gupta07} is that we focus on the virtual controller placement over fog by considering reliability and latency dimensions simultaneously. We show that the optimum virtual fog controller runs an estimator as a delay compensator, which does not appear in these papers. Further, the issue of reliability in our model is shifted from communication links to virtual fog controllers. 




\section{Fog Computing Architectures for Control-as-a-Service Applications \label{Section: Architectures}}

\par
In this section, we describe the properties of fog architectures that are important for placements of virtualized control services, and present the results of small-scale experiments with fog computing architectures.

\subsection{Fog Services: Heterogeneity and the Need for Auto-tuning}

\par 
Fog computing architectures are expected to include different physical links (wired, wireless, satellite), different extends of mobility of different nodes, and a wide range of differences in computing device capabilities~\cite{OpenFogReferenceArchitecture2017}.

\par
We expect the functionality in fog systems to be provided via \emph{service execution options with different performance parameters} (providing services for control applications can be referred to as creating \emph{Control-as-a-Service architectures}~\cite{Esen15_ControlAsService,Qi2016Design}). Cloud computing service providers are already offering a full range of service options that differ in the speed, cost, and complexity of execution~\cite{EC2InstanceTypes,AmazonLambda} -- the diversification that is likely to become more and more prominent in the future. 
Due to the inherently heterogenous nature of fog systems, 
we expect them to include a wider range of service execution options than the options provided in traditional cloud computing systems. In particular, we expect services in fog systems to be offered at a range of reliability options, starting from expensive high-availability services with ``five nines" uptime guarantees (i.e., 99.999\% availability, or the downtime of no more than 5.2 minutes per year)~\cite{Brewer2001Lessons}, to cheaper limited or frequently interrupted services provided by low-end nodes, including nodes with long sleep cycles 
and energy-harvesting-based intermittently powered nodes~\cite{Margolies2015EnHANTs,lucia2017Intermittent}.


\par 
Additionally, virtualized control functionality in fog systems will need to be \emph{placed, tuned, and moved around} automatically, without the inputs from the users. Existing commercial examples of automatic service placements in cloud computing include serverless computing mechanisms~\cite{GoogleCloudFunctions,AmazonLambda,AzureFunctions,OpenWhisk}, which use auto-provisioning (``autoscaling'') mechanisms to provide robustness to spikes in service request rates at the cost of additional latency~\cite{FowlerServerless}. In distributed heterogeneous fog computing settings, we will also require the ability to shift task execution point assignments to save energy, optimize deployment costs, and to free up constrained resources for critical tasks and services. Enabling the provision of such auto-optimizing virtual control services necessitates obtaining quantitative understanding of the tradeoffs between different control system design parameters that are not generally considered simultaneously.

\subsection{Latency and Reliability Tradeoffs: Small-scale Experiments}

\par
To better understand latency-reliability tradeoffs in fog computing systems, we conducted small-scale experiments with simple linear virtual controllers. 

\par 
The control service we implemented received, as inputs, a $2$-dimensional vector and a time index, looked up a corresponding $2$-by-$2$ matrix, performed matrix multiplication, and returned the results. This service implementation closely follows the controller operations we examine in this work. We executed the control application at different AWS Lambda \cite{AmazonLambda} cloud computing service points worldwide, at a Microsoft Azure serverless Functions computing point~\cite{AzureFunctions}, and on a local consumer-grade hardware device with an Intel Atom single-core 1.6GHz CPU. The local control service was implemented using a popular Flask \cite{FlaskFramework}
micro-service development framework over the built-in Flask HTTP server and over a Gunicorn WSGI HTTP Unix Server \cite{Gunicorn2017}.

\begin{table}[t]
\centering
  \caption{Latency of control service execution in different nodes.~\label{table:ExperimentalLatencies}}
  \vspace{-0.2cm}
 \begin{tabular}{|p{0.03\linewidth}|p{0.5\linewidth} | p{0.2\linewidth} | }
  \hline
  \# & Control service placement & Latency [secs] \\
 \hline
 1 & Local node & 0.06 \\ \hline
 2 & Microsoft Azure Functions US East & 0.08 \\ \hline
 3 & AWS Lambda US East (Virginia) & 0.5 \\  \hline
 4 & AWS Lambda US West (Seattle) & 0.8 \\ \hline
 5 & AWS Lambda Tokyo & 1.3 \\ \hline
 \end{tabular}
\end{table}

\par
Our small-scale experiments demonstrated the expected richness of virtualized fog control services in the latency-reliability space. Specifically, our response latency measurements, summarized in Table \ref{table:ExperimentalLatencies}, demonstrate that response times vary over a wide range. We observed a wide variety of reliability options in fog settings as well. Serverless AWS Lambda and Microsoft Azure Functions computing are provisioned on demand and hence can be seen as always available. 
Local control service, on the other hand, has only a finite number of control service processes deployed, and thus can handle only a fixed number of responses at the same time. Thus, as expected, we observed in our experiments with both default Flask server settings 
and with the Gunicorn server that when the instantiated processes are occupied, the responses can be deemed to be dropped in time-critical control applications due to large waiting times behind others. It is clear that these differences in reliability and latency 
are likely to be even more dramatic in {\em fully} heterogeneous fog computing architectures outlined above.

These experimental observations as well as the drone trajectory following problem provide the underlying motivation for the current paper to undertake a systematic study for delineating the direct and collateral effects of reliability and latency on virtual control services over fog.  

\section{Virtualized Control over Fog} \label{Section: System Model}
In this section, we will introduce the details of our system model, the definitions of the main concepts in relation to this model and the virtualized control problem over fog. 

\subsection{System Model}

\begin{figure}[!t]
\begin{center}
\includegraphics[width=1.0\linewidth]{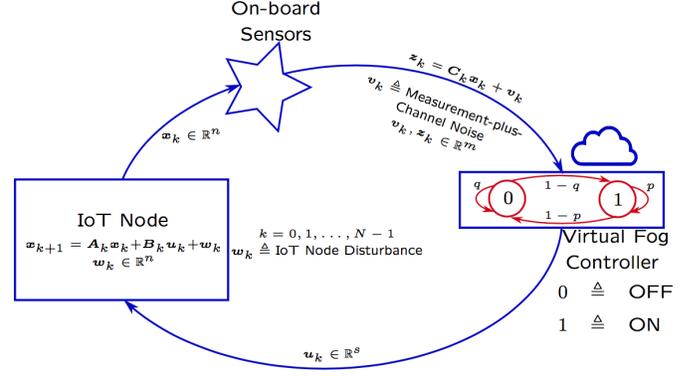}
\end{center}
\caption{Model for virtualized control over fog.} \label{Fig: System Model}
\end{figure}

We consider an IoT node (such as a UV, a robotic arm, etc.) whose dynamics evolve {\em linearly} in discrete-time according to
\begin{eqnarray}
\vec{x}_{k+1} = \vec{A}_k  \vec{x}_{k} + \vec{B}_k  \vec{u}_{k} + \vec{w}_k,  \label{Eqn: System Dynamics}
\end{eqnarray}
where $\vec{x}_{k} \in \R^n$ is the state vector, $\vec{u}_{k} \in \R^s$ is the control signal, $\vec{w}_k \in \R^n$ is the zero-mean random disturbance with covariance $\vec{W}_k$, and $\vec{A}_k \in \R^{n \times n}$ and $\vec{B}_k \in \R^{n \times s}$ are the system matrices that modulate the system states and control signals, respectively, for $k=0, \ldots, N-1$.\footnote{This paper does not consider how the time discretization is performed, which depends on the time-scale of change of the IoT process to be controlled as well as other several design degrees of freedom such as control quality and precision. For non-linear IoT node processes, it is assumed that Hartman-Grobman theorem holds and the system dynamics can be linearized around an equilibrium point \cite{Strogatz95}.}  The IoT node does not possess an on-board controller circuitry, and hence the controller functionality is virtualized on a fog endpoint, as shown in Fig. \ref{Fig: System Model}. The states are monitored by on-board sensors and are transmitted over a communication channel to the fog controller. The channel output is given by
\begin{eqnarray}
\vec{z}_k = \vec{C}_k \vec{x}_{k} + \vec{v}_k, \label{Eqn: Channel Equation}
\end{eqnarray}
where $\vec{C}_k \in \R^{m \times n}$ and $\vec{v}_k \in \R^m$ is the measurement-plus-channel noise.  We note the affinity of this model with the classical multiple-input multiple-output (MIMO) channel model \cite{Tse05}.\footnote{Our model can be considered to exemplify uncoded or low-complexity coded transmissions such as simple CRC schemes so that some measurement-plus-channel noise still remains in the received state measurements, which is a practical assumption for IoT nodes limited by computational resources and battery life.}  It is assumed that system disturbance and measurement-plus-channel noise vectors are independent among themselves as well as being independent over time. It is also assumed that reception of $\vec{z}_k$ marks a {\em request-for-control} and initiates the generation of a control signal at the fog endpoint.

The casual ordering of events to update IoT node states is the state measurement, forward delivery of measured states to the virtual fog controller, generation of control signals and backward delivery of control signals to the IoT node. 
When all these events happen in the same time-slot, we say that the fog controller is {\em perfectly matched} to the IoT node dynamics.  Otherwise, we say that it is {\em imperfectly matched}, in which case the control signals lag behind the IoT node state updates. To simplify exposition, we explain the rest of the system model for the case of perfect match, and relegate the details of the latter to Section \ref{Section: Control With Latency}.

We model the reliability issues observed in our fog computing experiments through a stylized Markov process having two states with transition probabilities
\begin{eqnarray*}
\PR{\tau_{k+1} = 1 \big| \tau_{k} = 1} = p
\end{eqnarray*}
and
\begin{eqnarray*}
\PR{\tau_{k+1} = 1 \big| \tau_{k} = 0} = 1-q,
\end{eqnarray*}
where $\tau_k$ is the internal state of the fog controller at time $k = 0, \ldots, N-1$. Here, the states $1$ (i.e., $\on$ state) and $0$ (i.e., $\off$ state) indicate the capacitated occupation status (due to multiple instantiated computing processes) of the fog endpoint for provisioning  the requested control service.

The control $\vec{u}_k$ depends on the available information at the fog controller by time $k$, which will be denoted as
\begin{eqnarray}
H_k = \brparen{\vec{z}_i: \tau_i = 1,  0 \leq i \leq k} \bigcup \brparen{\vec{u}_i: 0 \leq i \leq k-1}. \label{Eqn: System History}
\end{eqnarray}
We define $\hist_k$ to be the collection of all possible information sets available at time $k$. Note that the fog controller knows the previous control signals when generating $\vec{u}_k$ at time $k$, which is embodied in $H_k$. The class of dynamic control policies of interest to us in this paper is introduced in Definition \ref{Def: Control Policy} below.
\begin{definition} \label{Def: Control Policy}
A control policy is a sequence of functions $\vec{\pi} = \paren{\pi_0, \ldots, \pi_{N-1}}$ such that the $k$th component function $\pi_k: \hist_k \times \brparen{0, 1} \mapsto \R^s$ determines the control applied at time $k = 0, \ldots, N-1$, i.e., $\pi_k\paren{H_k, \tau_k} = \vec{u}_k$.
\end{definition}

We note that $\vec{\pi}$ is an on-line rule that observes the realizations of system history and determines the control signals to be applied based on these observations. We say that a control policy is {\em feasible} if $\pi_k\paren{H_k, 0} = 0$ for all $H_k \in \hist_k$ and $k=0, \ldots, N-1$.  That is, a feasible control policy does not output any control when the fog controller is at $\off$ state. An important subclass of control policies is that of memoryless ones, formally defined in Definition \ref{Def: Memoryless Control}.

\begin{definition} \label{Def: Memoryless Control}
A control policy is said to be {\em memoryless} if the control applied at time $k$ depends only on $\vec{z}_k$ and $\tau_k$ for $k=0, \ldots, N-1$.
\end{definition}

\subsection{Cost Minimization Problem}
As is standard in control systems \cite{Pappas15, Gupta07, Yuksel06, Yuksel13, Tatikonda04}, we rank the quality of virtualized control over fog by means of a quadratic cost function. In particular, we define per-stage-cost under control policy $\vec{\pi} = \paren{\pi_0, \ldots, \pi_{N-1}}$ at time $k$ as
\begin{eqnarray*}
g_{\pi_k}\paren{\vec{x}_k, H_k, \tau_k} = \vec{x}_k^\top \vec{Q}_k \vec{x}_k + \vec{u}_k^\top \vec{R}_k \vec{u}_k,
\end{eqnarray*}
where $\vec{u}_k$ is the control generated by $\vec{\pi}$ after having observed $H_k$, and $\vec{Q}_k$ is positive semidefinite and $\vec{R}_k$ positive definite for all $k$. Over a finite horizon of $N+1$ time-slots, the aggregate cost depends on three sources of randomness: (i) IoT node disturbance, (ii) channel noise and (iii) fog controller reliability. Hence, we write the total cost incurred over the time horizon of interest and averaged over the existing sources of randomness as

\begin{eqnarray}
\lefteqn{\cost{p}{q}} \hspace{8.0cm} \nonumber \\
\lefteqn{=\ESI{\vec{x}_N^\top \vec{Q}_N \vec{x}_N + \sum_{k=0}^{N-1} g_{\pi_k}\paren{\vec{x}_k, H_k, \tau_k}}{\paren{\vec{x}_0, \tau_0}},} \hspace{7.0cm} \label{Eqn: Total Cost}
\end{eqnarray}
where $\vec{x}_0$ and $\tau_0$ are the initial states and $\EW_{\paren{\vec{x}_0, \tau_0}}$ indicates expectation starting from these initial sates. Our aim is to minimize $\cost{p}{q}$ over the set of all feasible control policies $\Pi$, i.e.,
\begin{eqnarray}
\mincost{p}{q} = \inf_{\vec{\pi} \in \Pi} \cost{p}{q}. \label{Eqn: Optimum Cost}
\end{eqnarray}

We will solve the optimization problem in \eqref{Eqn: Optimum Cost} by utilizing the dynamic programming (DP) approach \cite{Bertsekas95}. To this end, we define cost-to-go functions as
\begin{eqnarray}
\lefteqn{\gocost{p}{q}{i}} \hspace{8.0cm} \nonumber \\
\lefteqn{= \ESI{\vec{x}_N^\top \vec{Q}_N \vec{x}_N + \sum_{k=i}^{N-1} g_{\pi_k}\paren{\vec{x}_k, H_k, \tau_k}}{\paren{\vec{x}_i, \tau_i}}.} \hspace{7.0cm} \label{Eqn: Cost-to-Go}
\end{eqnarray}
We note that $\gocost{p}{q}{i}$ depends on the $i$th tail policy $\vec{\pi}_i$  for $\vec{\pi}$, which is defined as $\vec{\pi}_i = \paren{\pi_i, \ldots, \pi_{N-1}}$.  The principle of optimality for DP states that the control policy $\vec{\pi}^\star$ having optimal tail policies $\vec{\pi}^{\star}_i = \paren{\pi_i^\star, \ldots, \pi_{N-1}^\star}$ for minimizing the cost-to-go function $\gocost{p}{q}{i}$ for any starting state $\vec{x}_i \in \R^n$ and $\tau_i \in \brparen{0, 1}$ for all $i \in \brparen{0, \ldots, N-1}$ solves the main DP problem in \eqref{Eqn: Optimum Cost}. We will utilize this fact for revealing the structure of the optimum control in both cases of perfectly and imperfectly matched fog controllers.

\section{Control over Fog With Perfect Match} \label{Section: Control Without Latency}
In this section, we will investigate the performance of virtual controllers with the focus of reliability being on the fog endpoint that is perfectly matched to the IoT node dynamics. Despite following the standard DP steps, our analysis in this section will be helpful to set the stage for the case of imperfectly matched fog controllers, which can only access to the delayed and intermittent state information from the IoT node.  We will consider the cases of fully observed and partially observed state information separately, starting with the former one below.

\subsection{Fully Observed State Information}

This is the case in which measurement and channel distortion can be ignored, i.e., $\vec{v}_k = \vec{0}$ and $\vec{C}_k$ is identity for all $k=0, \ldots, N-1$. Hence, the network does not mix-up state measurements while relaying them.  When the IoT node states are fully observed, it is then enough to focus on the memoryless policies \cite{Bertsekas95}, which is why we write the optimum policy below as a function of current states rather than the complete system history.  Theorem \ref{Thm: Min Cost Perfect State} establishes the recursive relationship for the optimum control policy $\vec{\pi}^\star$ and the min-cost performance under $\vec{\pi}^\star$.

\begin{theorem} \label{Thm: Min Cost Perfect State}
Assume $p = 1-q$ and the measured states can be {\em fully} observed with {\em perfect} match.  Then, the optimum control policy $\vec{\pi}^\star = \paren{\pi_0^\star, \ldots, \pi_{N-1}^\star}$ and $\mincost{p}{q}$ are given by
\begin{eqnarray}
\lefteqn{\mincost{p}{q} = \vec{x}_0^\top \paren{\vec{L}_0 - \vec{\Lambda}_0 \I{\tau_0 = 1}} \vec{x}_0} \hspace{8.0cm} \nonumber \\
\lefteqn{+\sum_{k=0}^{N-1} \tr{\vec{K}_{k+1} \vec{W}_k}} \label{Eqn: OptCost1}\hspace{3.0cm}
\end{eqnarray}
and
\begin{eqnarray}
\pi_k^\star\paren{\vec{x}_k, \tau_k} = -\vec{V}_k \vec{x}_k \I{\tau_k = 1}, \label{Eqn: Optimum Control1}
\end{eqnarray}
where $\vec{V}_k = \paren{\vec{R}_k + \vec{B}_k^\top \vec{K}_{k+1} \vec{B}_k}^{-1} \vec{B}_k^\top \vec{K}_{k+1} \vec{A}_k$ and the matrices $\vec{K}_k$, for $k=1, \ldots, N-1$, are given recursively as
\begin{eqnarray*}
\vec{K}_N &=& \vec{Q}_N \\
\vec{K}_k &=& \vec{L}_k - p \vec{\Lambda}_k \\
\vec{L}_k &=& \vec{Q}_k + \vec{A}_k^\top \vec{K}_{k+1} \vec{A}_k, \mbox{ and}\\
\vec{\Lambda}_k &=& \vec{A}_k^\top \vec{K}_{k+1} \vec{B}_k \vec{V}_k.
\end{eqnarray*}
\end{theorem}
\begin{IEEEproof}
Using the DP algorithm and bearing in mind the relation between feasible $\vec{u}_k$ and $\tau_k$, we write
\begin{eqnarray}
\mingocost{p}{q}{N} = \vec{x}_N^\top \vec{Q}_N \vec{x}_N \label{Eqn: Terminal Min Cost}
\end{eqnarray}
and
\begin{eqnarray}
\lefteqn{\mingocost{p}{q}{k} = \vec{x}_k^\top \vec{Q}_k \vec{x}_k + \min_{\vec{u}_{k} \in \R^s} \left\{ \vec{u}_{k}^\top \vec{R}_{k} \vec{u}_{k} \right.} \hspace{7.5cm} \nonumber \\
\lefteqn{\left. + \ESI{\mingocost{p}{q}{k+1} \big| \vec{u}_{k}}{\paren{\vec{x}_{k}, \tau_k}}\right\}} \hspace{5.2cm}
\end{eqnarray}
for the optimum cost-to-go expressions. We first consider the stage $N-1$. Using the IoT node process evolution in \eqref{Eqn: System Dynamics} and  minimizing the resulting quadratic form for $\tau_{N-1}=1$, one can obtain
\begin{eqnarray*}
\pi^\star_{N-1}\paren{\vec{x}_{N-1}, 1} = - \vec{V}_{N-1} \vec{x}_{N-1}
\end{eqnarray*}
and
\begin{eqnarray*}
\lefteqn{\mingocostON{p}{q}{N-1} = \vec{x}_{N-1}^\top \paren{\vec{L}_{N-1} - \vec{\Lambda}_{N-1}} \vec{x}_{N-1}} \hspace{8.0cm} \\
\lefteqn{+ \tr{\vec{K}_{N} \vec{W}_{N-1}}.} \hspace{2.1cm}
\end{eqnarray*}

On the other hand, no control is applied and the matrix $\vec{\Lambda}_{N-1}$ above disappears when $\tau_{N-1}=0$, which leads to the desired result for stage $N-1$ in Theorem \ref{Thm: Min Cost Perfect State}. For stage $N-2$, the same analysis holds, but one also needs to average over the Markov process transitions from $N-2$ to $N-1$. By considering the symmetry assumption, this leads to the optimum control given in \eqref{Eqn: Optimum Control1} and the minimum cost-to-go expression below
\begin{eqnarray*}
\lefteqn{\mingocost{p}{q}{N-2}} \hspace{8.0cm} \\
\lefteqn{= \vec{x}_{N-2}^\top \paren{\vec{L}_{N-2} - \vec{\Lambda}_{N-2} \I{\tau_{N-2} = 1}} \vec{x}_{N-2}} \hspace{5.8cm} \\
\lefteqn{+ \tr{\vec{K}_{N-1} \vec{W}_{N-2}} + \tr{\vec{K}_{N} \vec{W}_{N-1}}.} \hspace{5.8cm}
\end{eqnarray*}
Iterating similarly, one can complete the proof. \end{IEEEproof}

An appealing feature of the optimum control, given by Theorem \ref{Thm: Min Cost Perfect State}, for implementing a virtual controller at a fog endpoint is its linear structure, which is easy to implement over light-weight fog nodes. The effect of the fog endpoint reliability appears as a multiplicative coefficient in the definition of $\vec{K}_k$ matrices. In particular, as $p$ increases, the additive cost terms $\tr{\vec{K}_{k+1} \vec{W}_k}$, i.e., matrix traces, decrease and we start to have a smaller cost value in \eqref{Eqn: OptCost1}.\footnote{This follows from observing the fact that the matrices $\vec{K}_k$, $\vec{L}_k$ and $\vec{\Lambda}_k$ are positive semi-definite for all $k=0, \ldots, N-1$.} We note that this is also the same virtual control service structure implemented over AWS Lambda, Microsoft Azure, Flask HTTP and Gunicorn WSGI HTTP Unix servers in our experiments above. For the asymmetric case, a similar optimum control recursion can also be obtained after averaging over exponentially many sample $\on$-$\off$ scheduling tail-paths of the fog controller at each time $k = 0, \ldots, N-1$, which is however not computationally practical for a fog computing system. As a result, we provide performance upper and lower bounds as well as a low-complexity control policy achieving these bounds in the following theorem.

\begin{theorem} \label{Thm: Min Cost Bounds1}
Assume $p > 1-q$. Then,
\begin{eqnarray*}
\mincost{p}{q} \leq \mincost{1-q}{q}
\end{eqnarray*}
and
\begin{eqnarray*}
\mincost{p}{q} \geq \mincost{p}{1-p}.
\end{eqnarray*}
Moreover, the optimum control achieving $\mincost{1-q}{q}$ with {\em full} state information and {\em perfect} match also attains a performance in between these two bounds.
\end{theorem}
\begin{IEEEproof}
Intuitively, decreasing $p$ results in a less reliable fog computing system, which leads to the upper bound in the theorem. Similarly, decreasing $q$ leads to a more reliable fog computing system, which leads to the lower bound in the theorem. For the sake of exposition, the details are relegated to Appendix \ref{Appendix: Bounds 1}. \end{IEEEproof}

\subsection{Partially Observed State Information}
We now consider the case in which measured IoT node states can only be observed partially due to possible distortion in the measurement process and  communication environment. In this case, the optimum control policy derived in Theorem \ref{Thm: Min Cost Perfect State} loses its memoryless property and depends on the realizations of observed measurement history.  However, it is well-known that the information structure considered in this paper does not introduce any dual-effects, and control and estimation problems can be separated \cite{Bertsekas95, Tse74}, which leads to our next theorem.
\begin{theorem} \label{Thm: Min Cost Partial State}
Assume $p = 1-q$ and the measured states can be {\em partially} observed with {\em perfect} match.  Then, the optimum control policy $\vec{\pi}^\star = \paren{\pi_0^\star, \ldots, \pi_{N-1}^\star}$ and $\mincost{p}{q}$ are given by
\begin{eqnarray}
\lefteqn{\mincost{p}{q} = \vec{x}_0^\top \paren{\vec{L}_0 - \vec{\Lambda}_0 \I{\tau_0 = 1}} \vec{x}_0} \hspace{8.0cm} \nonumber \\ \lefteqn{+\sum_{k=0}^{N-1} \tr{\vec{K}_{k+1} \vec{W}_k} + \ESI{\vec{\epsilon}_0^\top \vec{\Lambda}_0 \I{\tau_0 = 1} \vec{\epsilon}_0}{\paren{\vec{x}_0, \tau_0}} } \hspace{6.5cm} \nonumber \\
\lefteqn{+ p \sum_{k=1}^{N-1} \ESI{\vec{\epsilon}_{k}^\top \vec{\Lambda}_{k} \vec{\epsilon}_{k}}{\paren{\vec{x}_0, \tau_0}}} \label{Eqn: OptCost2} \hspace{6.5cm}
\end{eqnarray}
and
\begin{eqnarray}
\pi_k^\star\paren{H_k, \tau_k} = -\vec{V}_k \ES{\vec{x}_k \big| H_k} \I{\tau_k = 1}, \label{Eqn: Optimum Control2}
\end{eqnarray}
where $\vec{\epsilon}_k = \vec{x}_k - \ES{\vec{x}_k \big| H_k}$ and the matrices $\vec{V}_k$, $\vec{K}_k$, $\vec{\Lambda}_k$ and $\vec{L}_k$ are as defined in Theorem \ref{Thm: Min Cost Perfect State}.
\end{theorem}
\begin{IEEEproof}
The proof follows from the existence of no dual-effects and the property of conditional expectations $\ES{\ES{X \big| \mathcal{F}_2} \big| \mathcal{F}_1} = \ES{X \big| \mathcal{F}_1}$ for any two nested $\sigma$-algebras $\mathcal{F}_1 \subseteq \mathcal{F}_2$ from \cite{Durrett96}.
\end{IEEEproof}

The first two terms in \eqref{Eqn: OptCost2} give the minimum cost attained in the former case of fully observed state information. On the other hand, the last two terms describe the effect of state estimation on the performance of virtual controller over the fog. Further, the structure of optimum control in \eqref{Eqn: Optimum Control2} is similar to the one in Theorem \ref{Thm: Min Cost Perfect State}, except the non-linear estimator $\ES{\vec{x}_k \big| H_k}$ that needs to be run at the fog endpoint, separately from the controller application. This can be an onerous and time-consuming task for light-weight fog nodes. Hence, it can replaced with its {\em linear} approximation through Kalman filtering in practical settings with some loss of optimality when the measurement-plus-channel noise is not Gaussian. The next theorem provides the analogous upper and lower bounds on the fog controller performance with partial state information.

\begin{theorem} \label{Thm: Min Cost Bounds2}
Assume $p > 1-q$. Then,
\begin{eqnarray*}
\mincost{p}{q} \leq \mincost{1-q}{q}
\end{eqnarray*}
and
\begin{eqnarray*}
\mincost{p}{q} \geq \mincost{p}{1-p}.
\end{eqnarray*}
Moreover, the optimum control achieving $\mincost{1-q}{q}$ with {\em partial} state information and {\em perfect} match also attains a performance in between these two bounds.
\end{theorem}
\begin{IEEEproof}
The proof follows from similar lines in Appendix \ref{Appendix: Bounds 1} by considering the same starting states and observed history for both systems in the inductive arguments, and hence is omitted to avoid repetition. \end{IEEEproof}

\section{Control over Fog With Imperfect Match} \label{Section: Control With Latency}
Despite being insightful when communication and computation latency can be ignored with respect to the IoT node state evolution dynamics, our analysis in Section \ref{Section: Control Without Latency}  cannot capture the wide variation of delay values measured in our virtual fog controller experiments, i.e., see Table \ref{table:ExperimentalLatencies}. Therefore, in this section, we will extend the basic model above to discover the effects of latency on the utility of fog controller placement along the cloud-to-things continuum. The augmented model is aimed to characterize the swiftness of control and timeliness of state measurements due to network delay between the IoT node and the fog endpoint.

Specifically, two types of delay are considered: (i) {\em forward} delay $M_F$ and (ii) backward delay $M_B$. $M_F$ is the delay incurred on the path from the IoT node to the fog endpoint at which the controller software runs, whereas $M_B$ is the delay in the reverse direction. To make the exposition simpler, we assume that any delay to compute optimum control is included in $M_B$ in the sequel.  The total delay incurred in both ways is then equal to $M = M_F + M_B$.\footnote{Here, we assume that these delays can either be reliably estimated or do not vary a lot around their means so that the discrete model for them becomes deterministic.} In this setup, any measurement about the IoT node state sent out at time $k$ arrives to the fog controller at time $k + M_F$, and a possible corrective control signal arrives back to the IoT node at time $k + M$. A potential  manifestation of this latency is a proportional decrease in the frequency of control actions arriving to the IoT node, which we model below by assuming delay-spreaded requests-for-control as in event-driven interactive control systems. We start our analysis with the case of fully observed state information.

\subsection{Fully Observed State Information}
This is the case in which transmitted state measurements can be fully observed by the fog controller with delay $M_F$. The next theorem establishes the optimum control rule and the min-cost performance under optimum control with fully observed but delayed state information.

\begin{theorem} \label{Thm: Min Cost Perfect State Imperfect Match}
Assume $p = 1-q$, $N \geq M\geq 1$ and the measured states can be {\em fully} observed with {\em imperfect} match. Define $c \defeq \frac{N-a}{M}$, where $a = N \mod M$ if $N$ is not an integer multiple of $M$, and $a = M$ otherwise.  Then, the optimum control policy $\vec{\pi}^\star = \paren{\pi_0^\star, \ldots, \pi_{N-1}^\star}$ and $\mincost{p}{q}$ are given by
\begin{eqnarray}
\lefteqn{\mincost{p}{q} = \vec{x}_0^\top \vec{L}_0 \vec{x}_0 +\sum_{k=0}^{N-1} \tr{\vec{K}_{k+1} \vec{W}_k}} \hspace{7.5cm} \nonumber \\
\lefteqn{+p \sum_{k=0}^{cM-1} \tr{\vec{P}_{k+1} \vec{W}_k}} \hspace{3.5cm} \label{Eqn: OptCost3}
\end{eqnarray}
and
\begin{eqnarray}
\lefteqn{\pi_k^\star\paren{\vec{\lambda}_{k}, \tau_{k-M_B}}} \hspace{8cm} \nonumber \\
\lefteqn{= \left\{\begin{IEEEeqnarraybox}[\relax][c]{l's} \vec{0} & if $k \neq 0 \mod M$\\
-\vec{V}_k \ES{\vec{x}_k \big|\vec{\lambda}_{k}} \I{\tau_{k-M_B} = 1} & if $k = 0 \mod M$ \end{IEEEeqnarraybox}\right.} \label{Eqn: Optimum Control3} \hspace{7.9cm}
\end{eqnarray}
for $k \in \brparen{1, \ldots, N-1}$, where $\vec{\lambda}_{k} = \paren{\vec{x}_{k-M}, \vec{u}_{k-M}}$ (for $k \geq M$), $\vec{V}_k = \paren{\vec{R}_k + \vec{B}_k^\top \vec{K}_{k+1} \vec{B}_k}^{-1} \vec{B}_k^\top \vec{K}_{k+1} \vec{A}_k$ and the matrices $\vec{K}_k$ and $\vec{P}_k$ are given recursively as
\begin{eqnarray*}
\vec{K}_N &=& \vec{Q}_N \\
\vec{K}_k &=& \left\{\begin{IEEEeqnarraybox}[\relax][c]{l's} \vec{L}_k & if $k \neq 0 \mod M$ \\
\vec{L}_k - p \vec{\Lambda}_k & if $k = 0 \mod M$
\end{IEEEeqnarraybox}\right. \\
\vec{L}_k &=& \vec{Q}_k + \vec{A}_k^\top \vec{K}_{k+1} \vec{A}_k, \mbox{ and} \\
\vec{\Lambda}_k &=& \vec{A}_k^\top \vec{K}_{k+1} \vec{B}_k \vec{V}_k
\end{eqnarray*}
for $k \in \brparen{0, \ldots, N-1}$, and
\begin{eqnarray*}
\vec{P}_{cM} &=& \vec{\Lambda}_{cM} \\
\vec{P}_k &=& \left\{\begin{IEEEeqnarraybox}[\relax][c]{l's} \vec{A}_k^\top \vec{P}_{k+1} \vec{A}_k & if $k \neq 0 \mod M$ \\
\vec{\Lambda}_k & if $k = 0 \mod M$
\end{IEEEeqnarraybox}\right.
\end{eqnarray*}
for $k \in \brparen{0, \ldots, cM-1}$.
\end{theorem}
\begin{IEEEproof}
We will provide the proof only for when $N$ is not an integer multiple of $M$.  The other case follows from observing that the final time a useful control arrives is $N-M$ and repeating the same steps below.

First, observe that no control arrives from time $cM + 1$ to $N$. Hence, we can write
\begin{eqnarray}
\mingocost{p}{q}{i} = \vec{x}_i^\top \vec{L}_i \vec{x}_i + \sum_{k=i}^{N-1} \tr{\vec{K}_{k+1}\vec{W}_k} \label{Eqn: Min Go Cost Delayed}
\end{eqnarray}
for $i \in \brparen{cM+1, \ldots, N}$. Second, we observe that control arrives to the IoT node for the final time at $cM$, generated by the fog controller at $cM - M_B$ based on full state observation $\vec{x}_{(c-1)M}$ and possible control $\vec{u}_{(c-1)M}$ at time $(c-1)M$. To this end,
the fog controller needs to solve
\begin{eqnarray*}
\min_{\vec{u} \in \R^s}\brparen{\vec{u}^\top \vec{R}_{i} \vec{u} + \ES{\ESI{\mingocost{p}{q}{i+1}\Big|\vec{u}}{\vec{x}_i} \Big| \vec{\lambda}_i}}
\end{eqnarray*}
for $i = cM$. Using \eqref{Eqn: Min Go Cost Delayed} and system linearity, the above minimization leads to
\begin{eqnarray*}
\pi_{cM}^\star\paren{\vec{\lambda}_{cM}, \tau_{cM-M_B}} = -\vec{V}_{cM} \ES{\vec{x}_{cM} \big|\vec{\lambda}_{cM}} \I{\tau_{cM-M_B} = 1}
\end{eqnarray*}
and, with a slight abuse of notation\footnote{$\mingocost{p}{q}{cM}$ actually depends on $\tau_{cM-M_B}$ but we have chosen to use above notation for the sake notational consistency.},
\begin{eqnarray*}
\lefteqn{\mingocost{p}{q}{cM} = \vec{x}_{cM}^\top \paren{\vec{L}_{cM} - \vec{\Lambda}_{cM} \I{\tau_{cM-M_B}=1}} \vec{x}_{cM}} \hspace{8cm} \\
\lefteqn{+\sum_{k=cM}^{N-1} \tr{\vec{K}_{k+1} \vec{W}_k} + \vec{\epsilon}_{cM}^\top \vec{P}_{cM} \I{\tau_{cM-M_B}=1} \vec{\epsilon}_{cM},} \hspace{7.0cm}
\end{eqnarray*}
where $\vec{\epsilon}_{cM} = \vec{x}_{cM} - \ES{\vec{x}_{cM} \big| \vec{\lambda}_{cM}}$. Now observing that no control arrives at time $cM-1$ and using the symmetry assumption as well as system linearity, we have
\begin{eqnarray}
\lefteqn{\mingocost{p}{q}{i} = \vec{x}_i^\top \vec{L}_i \vec{x}_i + \sum_{k=i}^{N-1} \tr{\vec{K}_{k+1}\vec{W}_k}} \hspace{8cm} \nonumber \\
\lefteqn{+p\tr{\vec{P}_{i+1} \vec{W}_i} + p \vec{\epsilon}_i^\top \vec{P}_i \vec{\epsilon}_i} \hspace{5.8cm} \nonumber
\end{eqnarray}
for $i = cM - 1$ and $\vec{\epsilon}_{cM-1} = \vec{x}_{cM-1} - \ES{\vec{x}_{cM-1} \big| \vec{\lambda}_{cM}}$, which is the {\em residual} error from estimation at time $cM$.

In order to complete the proof, we repeat the same steps until $(c-1)M$. Observing that the residual error from estimation at $cM$ is equal to $\vec{w}_{(c-1)M}$ at $(c-1)M$ and taking the new estimation error appearing due to potential control to be applied at $(c-1)M$ into account, we arrive at
\begin{eqnarray}
\lefteqn{\mingocost{p}{q}{i}} \hspace{8cm} \nonumber \\
\lefteqn{= \vec{x}_i^\top \paren{\vec{L}_i - \vec{\Lambda}_{i} \I{\tau_{i-M_B}=1}} \vec{x}_i + \sum_{k=i}^{N-1} \tr{\vec{K}_{k+1}\vec{W}_k}} \hspace{7.2cm} \nonumber \\
\lefteqn{+p \sum_{k=i}^{cM-1} \tr{\vec{P}_{k+1} \vec{W}_k} +  \vec{\epsilon}_i^\top \vec{P}_i \I{\tau_{i-M_B}=1} \vec{\epsilon}_i,} \hspace{7.2cm} \nonumber
\end{eqnarray}
where $\vec{\epsilon}_{i} = \vec{x}_{i} - \ES{\vec{x}_{i} \big| \vec{\lambda}_{i}}$ for $i = (c-1)M$. Observing the emerging structure, and iterating similarly first from $(c-1)M$ to $M$ and then from $M$ to $0$, one can complete the proof. \end{IEEEproof}

There are several important structural features of the optimum control and the resulting minimum cost appearing in Theorem \ref{Thm: Min Cost Perfect State Imperfect Match}. First, the linearity property is preserved even with imperfect match between the IoT node process and the fog controller placement. This is important for practical low-complexity implementations of the controller application at the fog endpoint. Second, the separation principle still holds in the case of imperfect match. An important ramification of this observation is that an estimator needs to be run separately from the controller software at the fog endpoint, which can be interpreted as a delay compensator.  However, when compared to the estimation problem with partial state observation in \eqref{Eqn: Optimum Control2}, this estimator is linear and easy to implement due to linear evolution of IoT node states in \eqref{Eqn: System Dynamics}. Third, the frequency of corrective control signals arriving to the IoT node is decreased in proportion to $M$. This is due to the aggregate delay over the fog network that spreads transmitted measurements and control signals. Finally, the last summation in \eqref{Eqn: OptCost3} represents the collateral impact of latency on the min-cost performance due to delayed state measurements, which would disappear for an hypothetical perfect estimator.\footnote{$\mincost{p}{q}$ in \eqref{Eqn: OptCost3} does not depend on $\tau_0$ since the system forgets about the starting fog node state after one unit delay due to symmetry assumption.}

{\bf Virtual Controller Placement:} Equations \eqref{Eqn: OptCost1} and \eqref{Eqn: OptCost3} in Theorems \ref{Thm: Min Cost Perfect State} and \ref{Thm: Min Cost Perfect State Imperfect Match} provide a pair of quantative performance values for fog network designers to decide about where to place virtual controller software by considering inherent service grade and latency characteristics of available fog endpoints when measurement and channel distortion can be ignored. The next theorem provides the analogous upper and lower bounds on the fog controller performance, as is done in Section \ref{Section: Control Without Latency}.

\begin{theorem} \label{Thm: Min Cost Bounds3}
Assume $p > 1-q$. Then,
\begin{eqnarray*}
\mincost{p}{q} \leq \mincost{1-q}{q}
\end{eqnarray*}
and
\begin{eqnarray*}
\mincost{p}{q} \geq \mincost{p}{1-p}.
\end{eqnarray*}
Moreover, the optimum control achieving $\mincost{1-q}{q}$ with {\em full} state information and {\em imperfect} match also attains a performance in between these two bounds.
\end{theorem}
\begin{IEEEproof}
The proof follows from similar lines in Appendix \ref{Appendix: Bounds 1} by observing that the inductive hypothesis holds for $k= cM, \ldots, N$, and considering the same starting states and observed measurements in the inductive arguments for any $k < cM$. It is omitted to avoid repetition. \end{IEEEproof}

\subsection{Partially Observed State Information}
Finally, we analyze the case with partially observed states and imperfect match between the IoT node and fog controller. Updating the definition of system history in \eqref{Eqn: System History} in an obvious way to include only delayed measurements and repeating the same steps above, we obtain the following theorems.

\begin{theorem} \label{Thm: Min Cost Partial State Imperfect Match}
Assume $p = 1-q$, $N \geq M\geq 1$ and the measured states can be {\em partially} observed with {\em imperfect} match. Define $c \defeq \frac{N-a}{M}$, where $a = N \mod M$ if $N$ is not an integer multiple of $M$, and $a = M$ otherwise.  Then, the optimum control policy $\vec{\pi}^\star = \paren{\pi_0^\star, \ldots, \pi_{N-1}^\star}$ and $\mincost{p}{q}$ are given by
\begin{eqnarray}
\lefteqn{\mincost{p}{q}} \hspace{8cm} \nonumber \\
\lefteqn{= \vec{x}_0^\top \vec{L}_0 \vec{x}_0 +\sum_{k=0}^{N-1} \tr{\vec{K}_{k+1} \vec{W}_k} +p \sum_{k=0}^{cM-1} \tr{\vec{P}_{k+1} \vec{W}_k} } \hspace{7.8cm} \nonumber \\
\lefteqn{+ p\sum_{k=1}^{c-1} \ESI{\vec{\epsilon}^\top_{kM} \vec{P}_{kM} \vec{\epsilon}_{kM}}{\paren{\vec{x}_0, \tau_0}}} \hspace{7.8cm} \label{Eqn: OptCost4}
\end{eqnarray}
and
\begin{eqnarray}
\lefteqn{\pi_k^\star\paren{H_{k}, \tau_{k-M_B}}} \hspace{8cm} \nonumber \\
\lefteqn{= \left\{\begin{IEEEeqnarraybox}[\relax][c]{l's} \vec{0} & if $k \neq 0 \mod M$\\
-\vec{V}_k \ES{\vec{x}_k \big|H_{k}} \I{\tau_{k-M_B} = 1} & if $k = 0 \mod M$ \end{IEEEeqnarraybox}\right.} \label{Eqn: Optimum Control4} \hspace{7.9cm}
\end{eqnarray}
for $k \in \brparen{1, \ldots, N-1}$, where the matrices $\vec{V}_k$, $\vec{K}_k$, $\vec{\Lambda}_k$, $\vec{L}_k$ and $\vec{P}_k$ are as defined in Theorem \ref{Thm: Min Cost Perfect State Imperfect Match}, and $\vec{\epsilon}_{kM} = \vec{x}_{kM} - \ES{\vec{x}_{kM} \big| H_{(k+1)M}}$ for $k \in \brparen{1,\ldots, c-1}$.
\end{theorem}
\begin{IEEEproof}
The proof follows from the same steps in the proof of Theorem \ref{Thm: Min Cost Perfect State Imperfect Match}, existence of no dual-effects and the property of conditional expectations $\ES{\ES{X \big| \mathcal{F}_2} \big| \mathcal{F}_1} = \ES{X \big| \mathcal{F}_1}$ for any two nested $\sigma$-algebras $\mathcal{F}_1 \subseteq \mathcal{F}_2$ from \cite{Durrett96}.
\end{IEEEproof}

The last summation in \eqref{Eqn: OptCost4} represents the error terms arising from the uncertainty that cannot be resolved via partial state observations. As in the case of perfect match with partial state observations, the estimation in \eqref{Eqn: Optimum Control4} is non-linear, which can be replaced with its linear version through practical Kalman filtering implementation at the expense of some loss of optimality.

{\bf Virtual Controller Placement:} Equations \eqref{Eqn: OptCost2} and \eqref{Eqn: OptCost4} in Theorems \ref{Thm: Min Cost Partial State} and \ref{Thm: Min Cost Partial State Imperfect Match} provide a pair of quantative performance values for fog network designers to decide about where to place virtual controller software by considering inherent service grade and latency characteristics of available fog endpoints when measurement and channel distortion cannot be ignored.

\begin{theorem} \label{Thm: Min Cost Bounds4}
Assume $p > 1-q$. Then,
\begin{eqnarray*}
\mincost{p}{q} \leq \mincost{1-q}{q}
\end{eqnarray*}
and
\begin{eqnarray*}
\mincost{p}{q} \geq \mincost{p}{1-p}.
\end{eqnarray*}
Moreover, the optimum control achieving $\mincost{1-q}{q}$ with {\em partial} state information and {\em imperfect} match also attains a performance in between these two bounds.
\end{theorem}
\begin{IEEEproof}
The proof follows from similar lines in Appendix \ref{Appendix: Bounds 1} by observing that the inductive hypothesis holds for $k= cM, \ldots, N$, and considering the same starting states and observed system history in the inductive arguments for any $k < cM$. It is omitted to avoid repetition. \end{IEEEproof}

\section{UV Trajectory Control over Fog} \label{Section: Numerical Results}
In this section, we illustrate the utility of our analytical results derived above in the context of trajectory tracking problems for UVs. While our work applies to all linear systems with quadratic cost, we note that our model maps well to practical drone trajectory tracking control problems. For example, modern high-end DJI quadcopters~\cite{DJIMatrice2017} can be programmed with a sequence of way-points, and can receive velocity adjustment signals. Our virtual controller model can also be used to dictate different levels of trajectory tracking precision in different parts of the trajectory, which is important for practical drone settings where more precision is required close to stationary obstacles, other drones, and no-fly zones than in unrestricted air spaces.

\subsection{State-space Representation for UV Control}

We consider a planar motion at some certain altitude determined through a sequence of way-points $\brparen{\bar{\vecbold{x}}_k}_{k=0}^N \subset \R^2$. $\bar{\vecbold{x}}_k$ represents the desired position of the UV at time $k \Delta t$. Here, $\Delta t$ is our basic discrete-time unit to communicate location information (obtained through GPS sensors) and control signals between the fog server and UV.\footnote{The model can be extended to include time-varying discrete-time units. However, we do not pursue this direction in the current paper for the sake of notational simplicity.}

The task of the fog controller is to provide velocity adjustment signals represented by $\vecbold{u}_k$ (measured in meters per second) for $k=0, \ldots, N-1$ to determine the velocity of the UV from time $k \Delta t$ to $\paren{k+1} \Delta t$. Succinctly, the state update equation can be given as
\begin{eqnarray}
\left(\begin{array}{c} \vecbold{x}_{k+1} \\ \vecbold{v}_{k+1} \end{array}\right) = \left(\begin{array}{c}\vecbold{x}_{k} + \Delta t \paren{\vecbold{v}_k + \vecbold{u}_k} \\ \vecbold{v}_{k} + \vecbold{u}_k \end{array}\right) + \left(\begin{array}{c}\vecbold{w}_{k}^x \\ \vecbold{w}_k^v\end{array}\right), \label{Eqn: Autonomous Vehicle 1}
\end{eqnarray}
where $\vecbold{w}_{k}^x$ and $\vecbold{w}_k^v$ are random (possibly correlated) disturbances affecting the location and velocity of the UV due to environmental conditions such as wind and rain, respectively. After some manipulations, \eqref{Eqn: Autonomous Vehicle 1} can be written as 

\begin{eqnarray}
\left(\begin{array}{c} \vecbold{e}_{k+1} \\ \vecbold{v}_{k+1} \end{array}\right) = \vecbold{A}_k \left(\begin{array}{c} \vecbold{e}_{k} \\ \vecbold{v}_{k} \end{array}\right) + \vecbold{B}_k \vecbold{u}_k + \left(\begin{array}{c}\vecbold{w}_{k}^x + \bar{\vecbold{x}}_{k} - \bar{\vecbold{x}}_{k+1} \\ \vecbold{w}_k^v\end{array}\right) \nonumber
\end{eqnarray}
where $\vecbold{e}_{k} = \vecbold{x}_{k} - \bar{\vecbold{x}}_{k}$, $\vecbold{A}_k = \left(\begin{array}{cc}\vecbold{I}_{2} & \Delta t \vecbold{I}_2 \\ \vecbold{0} & \vecbold{I}_2\end{array}\right)$ and $\vecbold{B}_k = \left(\begin{array}{c}\Delta t \vecbold{I}_{2} \\ \vecbold{I}_{2} \end{array}\right)$ for $k=0, \ldots, N-1$.\footnote{More explicitly, the randomly varying velocity process of the UV can be written as $\vecbold{V}(t) = \vecbold{v}_k + \vecbold{u}_k + \vecbold{W}(t)$ for $k \Delta t < t \leq (k+1)\Delta t $, where $\vecbold{W}(t)$ is the stochastic disturbance process inflicting the UV motion.  Then, $\vecbold{w}_{k}^x$ is the integral of $\vecbold{W}(t)$ from $k\Delta t$ to $(k+1)\Delta t$, and $\vecbold{w}_k^v$ is the value of $\vecbold{W}(t)$ at time $(k+1)\Delta t$.} This is similar to the linear IoT node model in Section \ref{Section: System Model}, with drift terms providing desired trajectory information. Hence, the optimum control will direct the UV along the desired path while trying to minimize drifted state measurements. To this end, we minimize the following cost $J = \sum_{k=0}^N \ES{\abs{\vecbold{e}_k}^2 + \alpha\paren {\abs{\vecbold{v}_k}^2 + \abs{\vecbold{u}_k}^2}}$ in an attempt to balance trajectory deviations and energy expenditure, which leads to $\vecbold{Q}_k = \left(\begin{array}{cc}\vecbold{I}_{2} & \vecbold{0} \\ \vecbold{0} & \alpha \vecbold{I}_2\end{array}\right)$, $\vecbold{R}_k = \alpha \vecbold{I}_2$ and $\alpha \geq 0$ being a design parameter. 
In this cost expression, we note that the terms containing $\abs{\vecbold{e}_k}^2$ measure the total deviation from the desired trajectory, whereas the remaining terms act as a proxy for the total energy spent to follow it.

\subsection{Trajectory Tracking Performance}

We consider a target located at a given position in space.  The objective of the UV is to approach the target (represented by a red cross in the figures below), make a circle around it and then turn back to its starting position.  The disturbances in location and velocity are jointly Gaussian with mean zero and the covariance matrix $\vecbold{\Sigma} = \left(\begin{array}{cc}\sigma^2_x \vecbold{I}_2 & \rho \vecbold{I}_2 \\\rho \vecbold{I}_2 & \sigma^2_v \vecbold{I}_2\end{array}\right)$.


\begin{figure*}[!t]
\begin{minipage}[t]{20cm}
\begin{center}
\hspace{-1.75cm}
\includegraphics[scale=0.23]{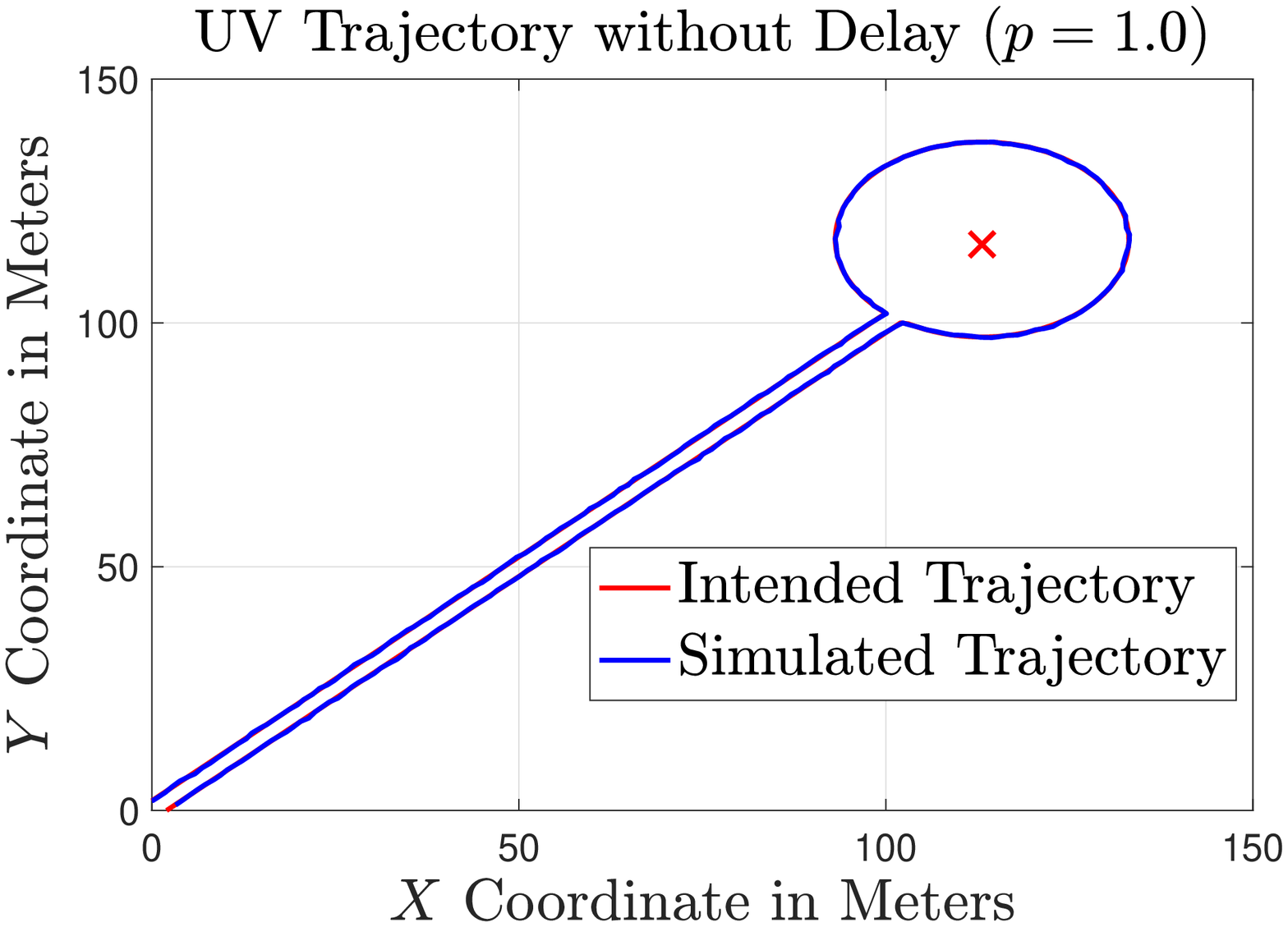}
\hspace{-0.5cm}
\includegraphics[scale=0.23]{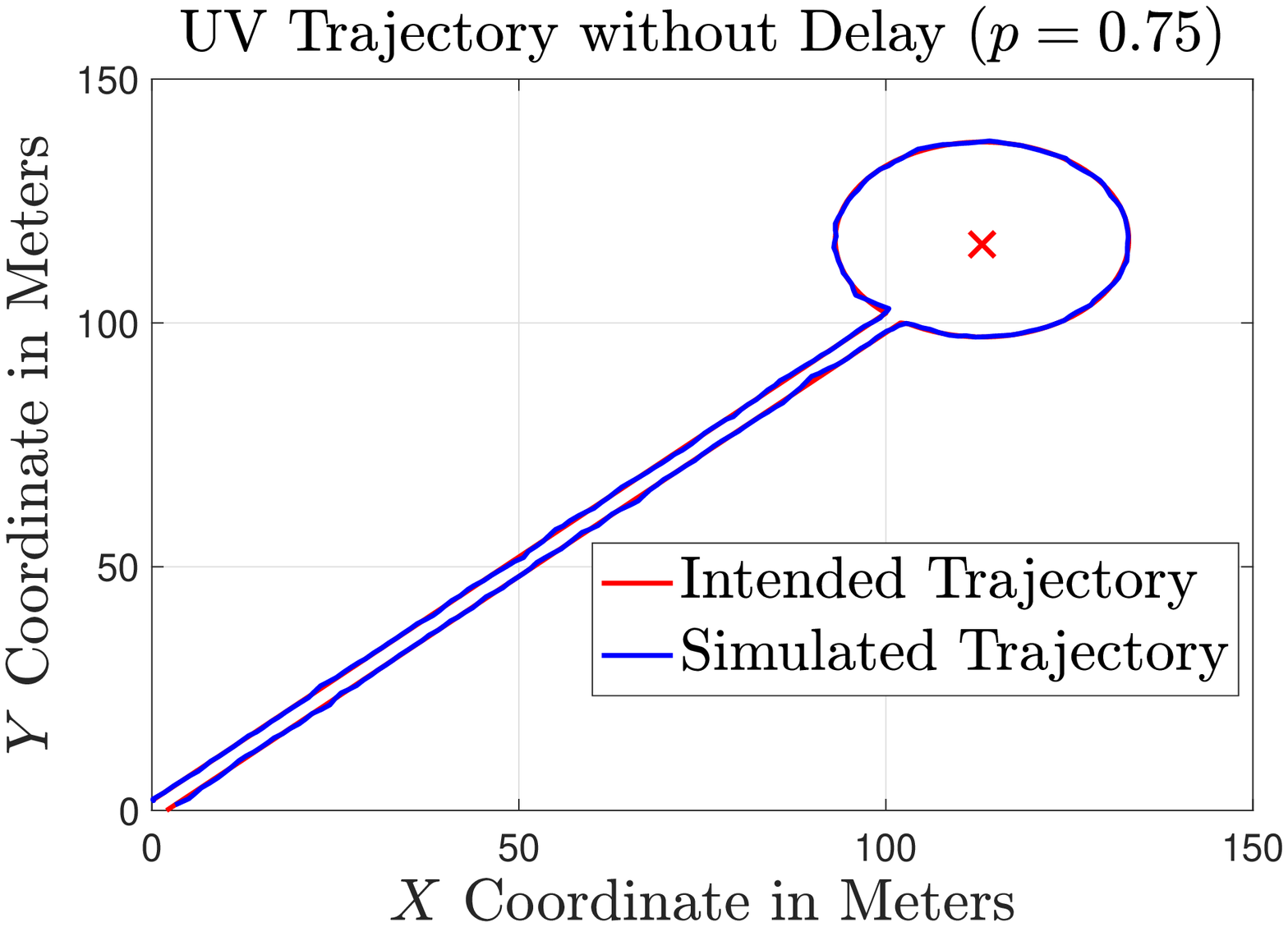}
\hspace{-0.5cm}
\includegraphics[scale=0.23]{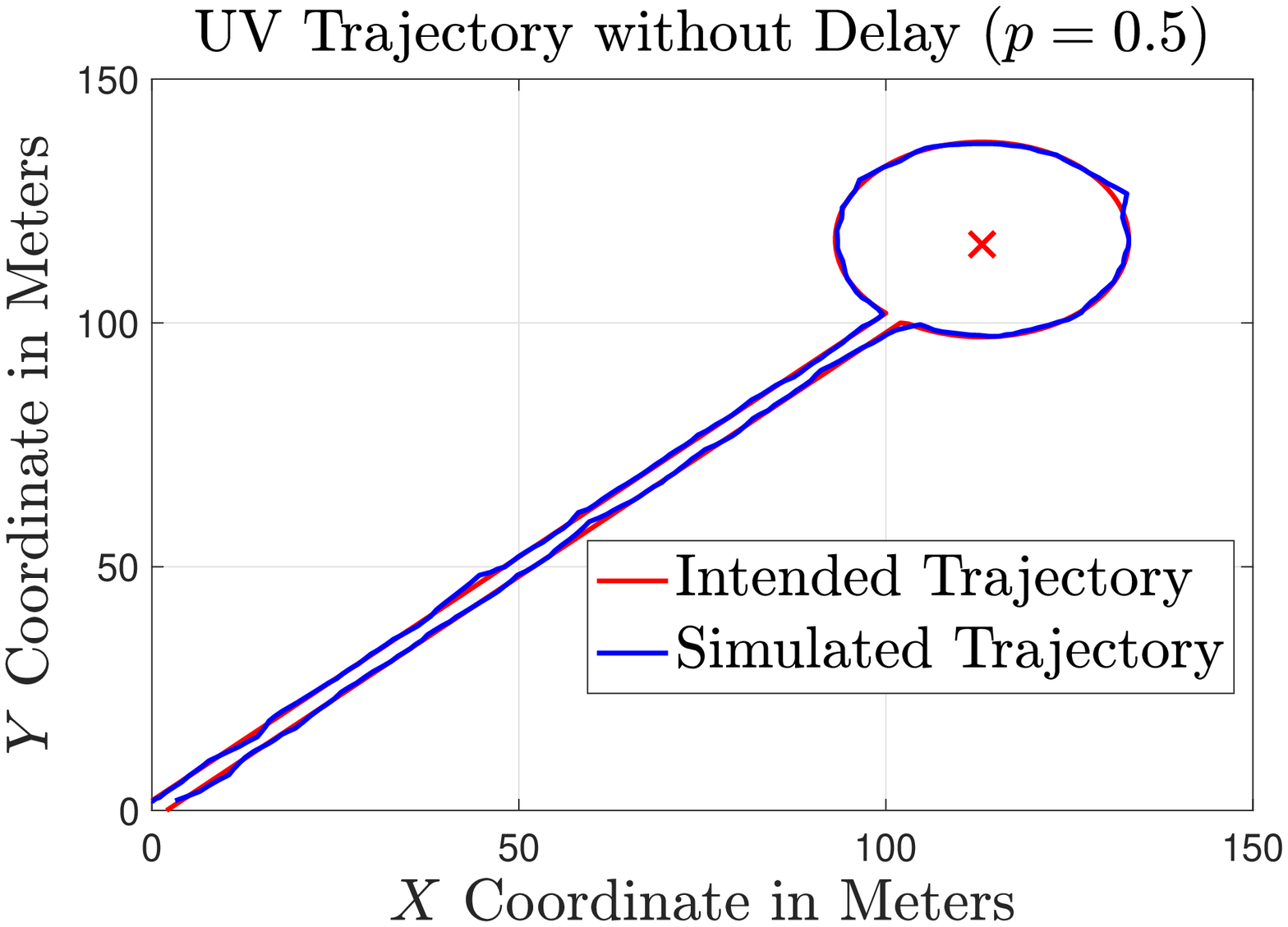}
\hspace{-0.5cm}
\includegraphics[scale=0.23]{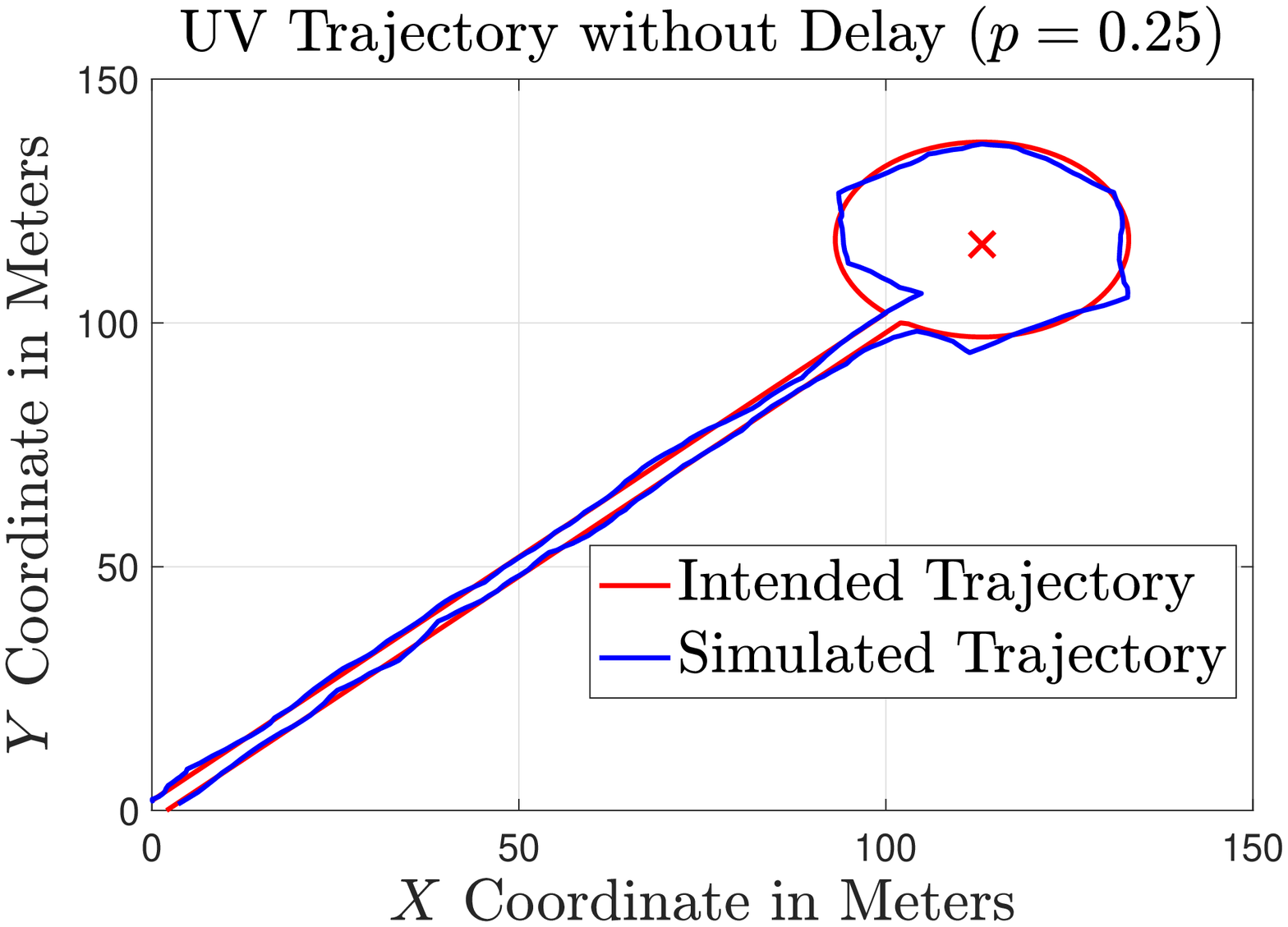}
\end{center}
\end{minipage}
\\
\begin{minipage}[t]{20cm}
\begin{center}
\hspace{-1.75cm}
\includegraphics[scale=0.23]{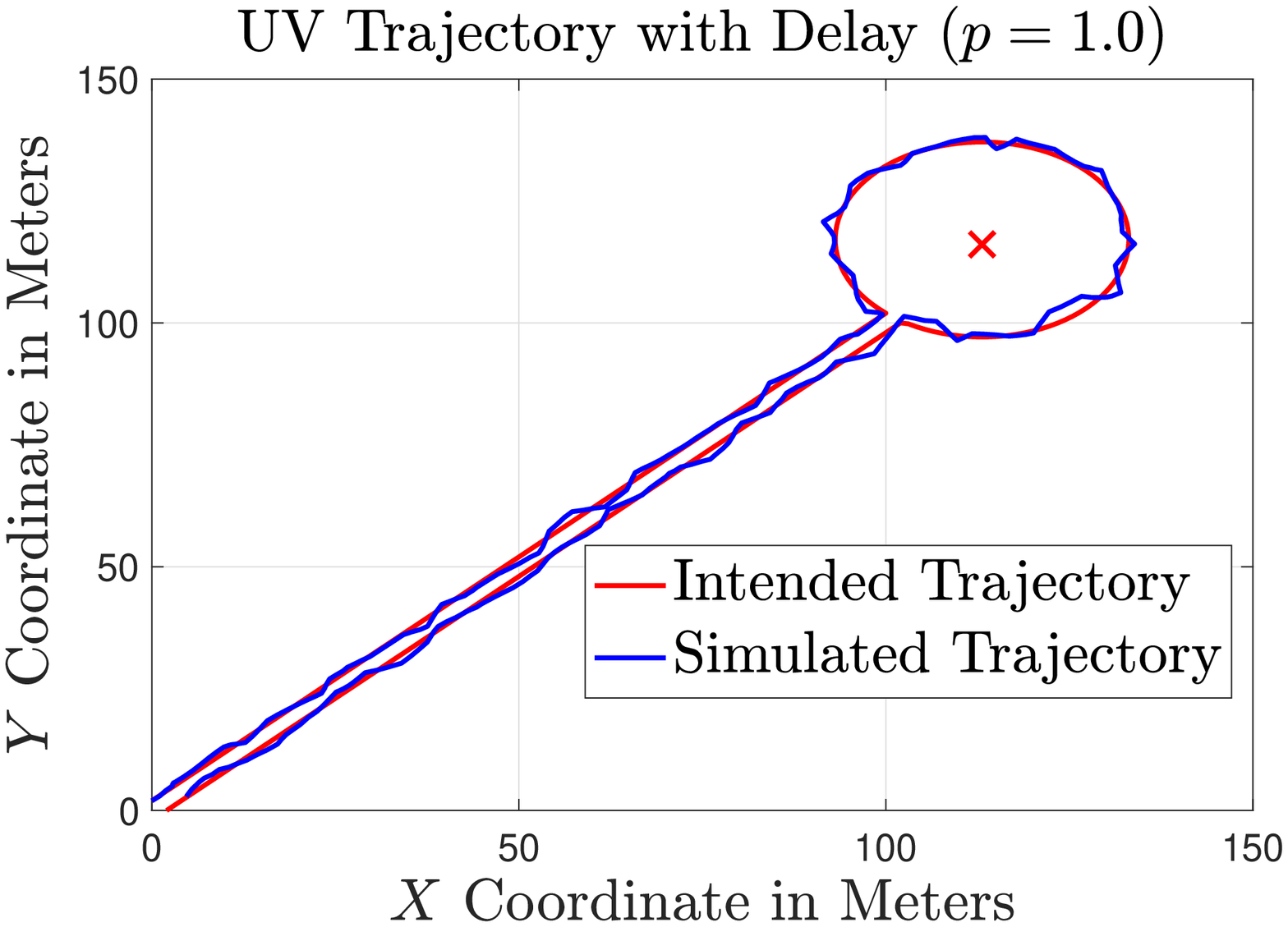}
\hspace{-0.5cm}
\includegraphics[scale=0.23]{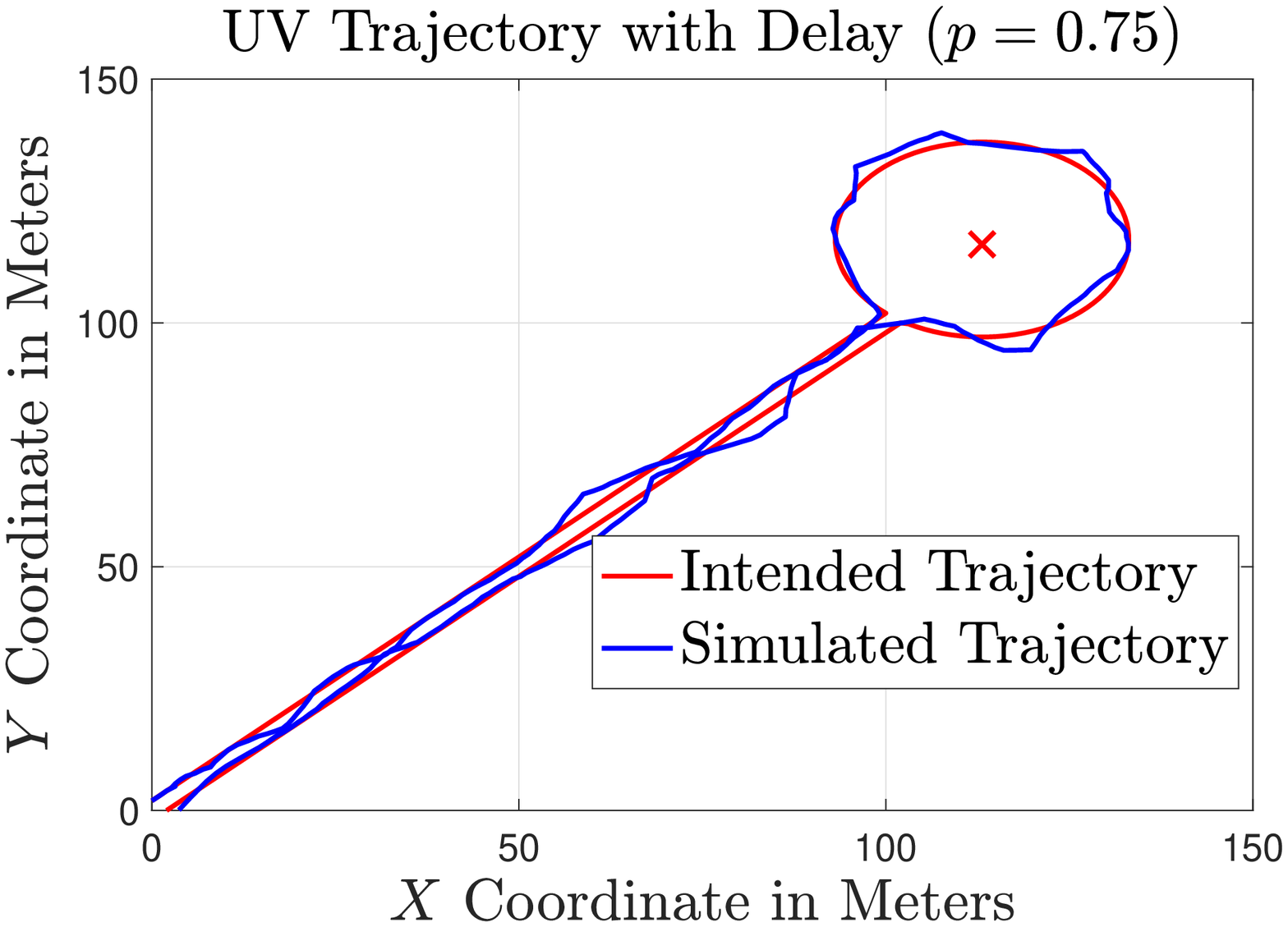}
\hspace{-0.5cm}
\includegraphics[scale=0.23]{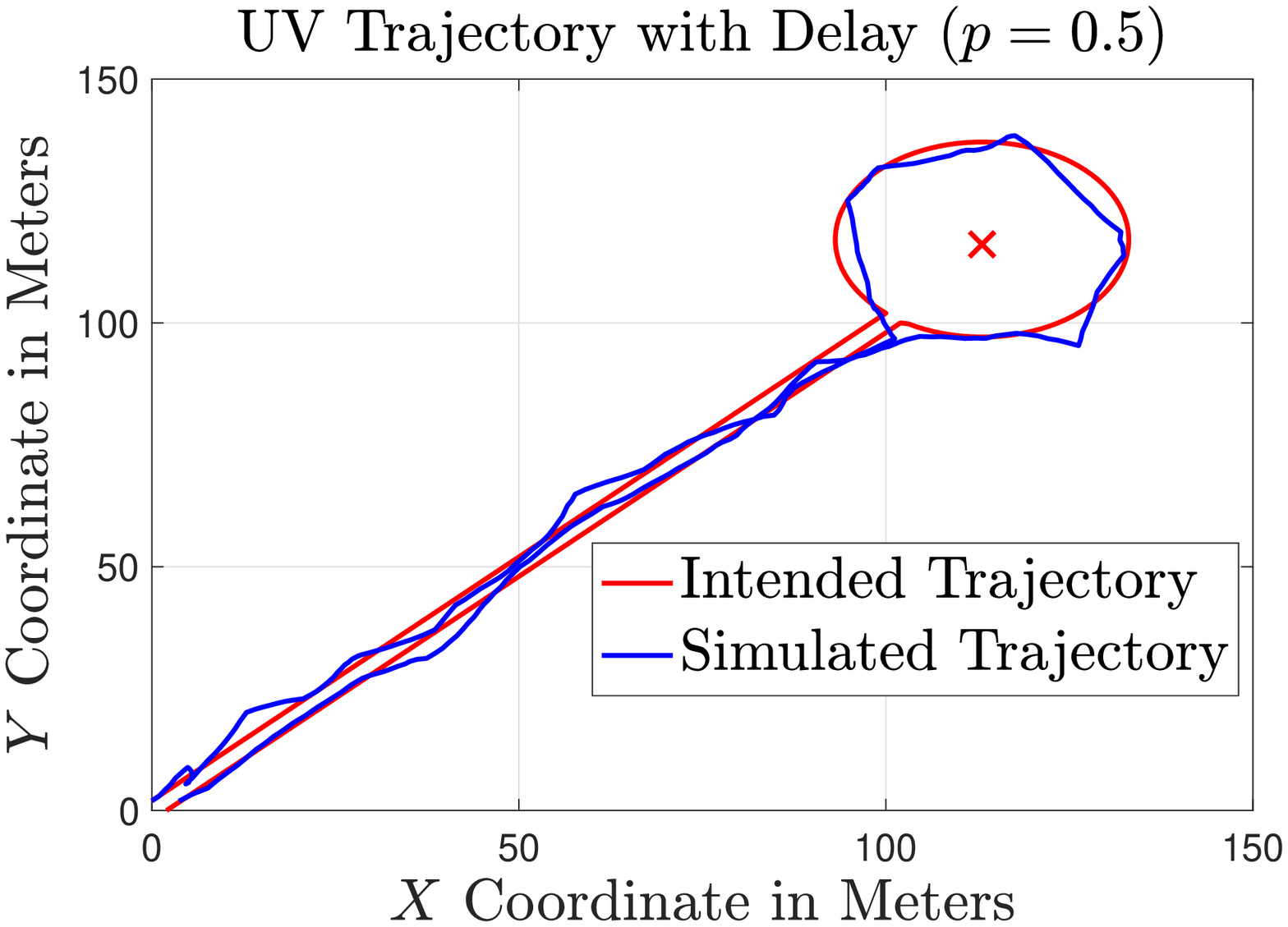}
\hspace{-0.5cm}
\includegraphics[scale=0.23]{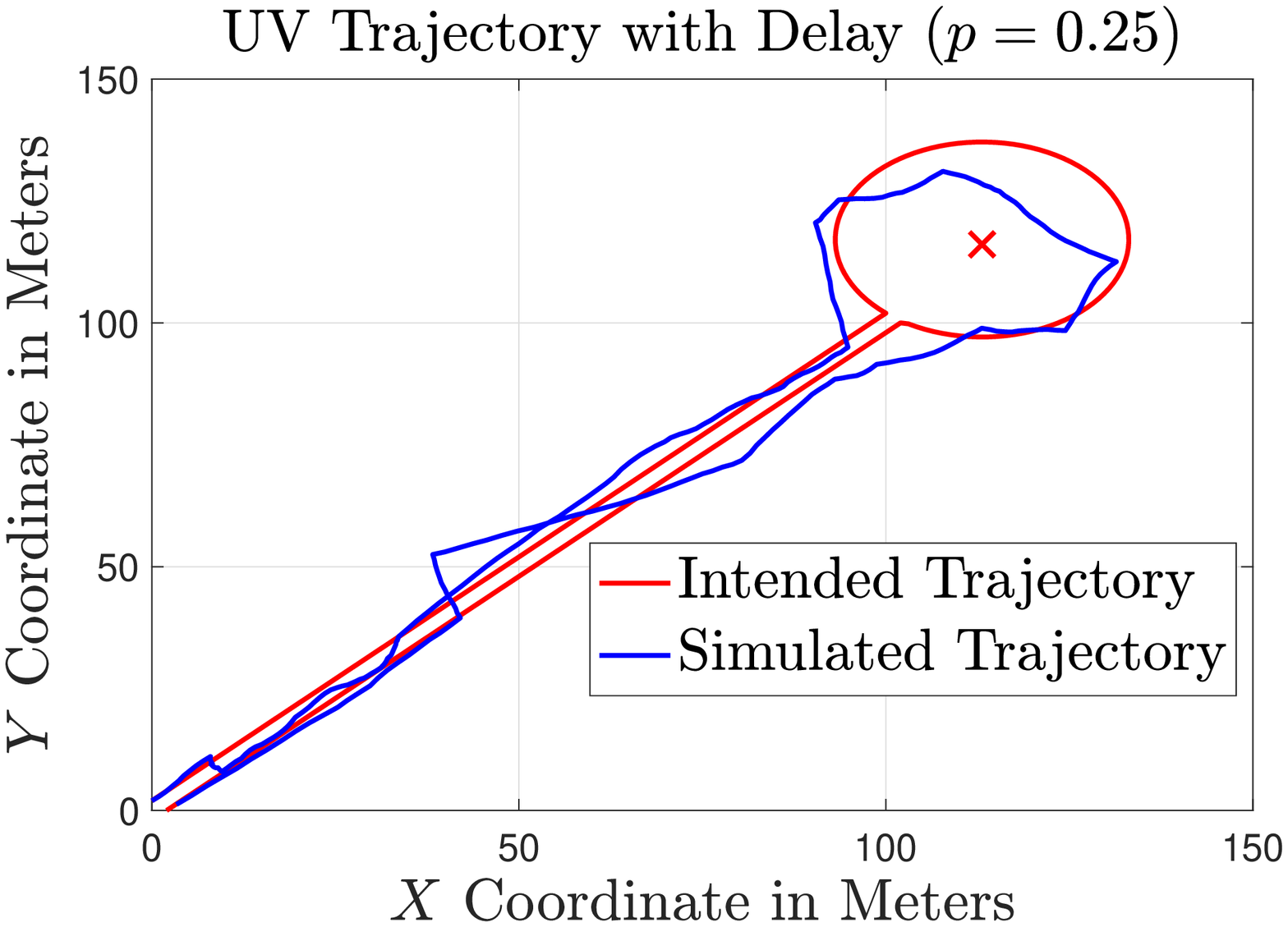}
\end{center}
\end{minipage}
\caption{Effect of reliability on the trajectory tracking performance of a UV fog controller. ($\alpha = 0.1, \sigma_x = 0.1$, $\sigma_v = 0.1$ and $\rho = \frac{\sigma_x \sigma_v}{2}$ for all figures.  The upper figures are for the case without delay, whereas the lower figures are for the case with delay $M = 3 \Delta t$.)}  \label{Fig: Figure 2}
\end{figure*}

In Fig. \ref{Fig: Figure 2}, we show the realizations of the vehicle trajectory for different values of the fog controller reliability parameter $p$.  In this figure, we set the system parameters $\alpha, \sigma_x$ and $\sigma_y$ to $0.1$ in order to delineate the trajectory tracking performance of the fog controller with respect to its reliability and delay.  We chose $\rho$ to have a correlation coefficient of $0.5$ in all cases.  In the upper figures, we observe that our optimum fog controller without delay is able to stabilize the UV around the predefined desired trajectory.  As expected, the performance of the fog controller in tracking the desired trajectory improves when $p$ increases. In particular, the UV almost perfectly tracks the desired trajectory for values of $p$ above $0.5$, which can used, for example, as the minimum level of virtual controller reliability for control services to be provisioned over fog in this particular case.

In the lower figures in Fig. \ref{Fig: Figure 2}, on the other hand, we illustrate the performance of the fog controller with delay $M = 3 \Delta t$.  Even with a very reliable fog controller, we observe substantial negative effects of delay.  In particular, the fog controller loses its ability to guide the UV around the desired trajectory starting from $p=0.5$ and downwards even with small disturbance.  These observations perfectly illustrate the potential of our analytical expressions to guide the design and engineering efforts to offer control-as-a-service over fog by considering reliability and delay constraints.

\begin{figure*}[!t]
\begin{minipage}[t]{20cm}
\begin{center}
\hspace{-1.75cm}
\includegraphics[scale=0.23]{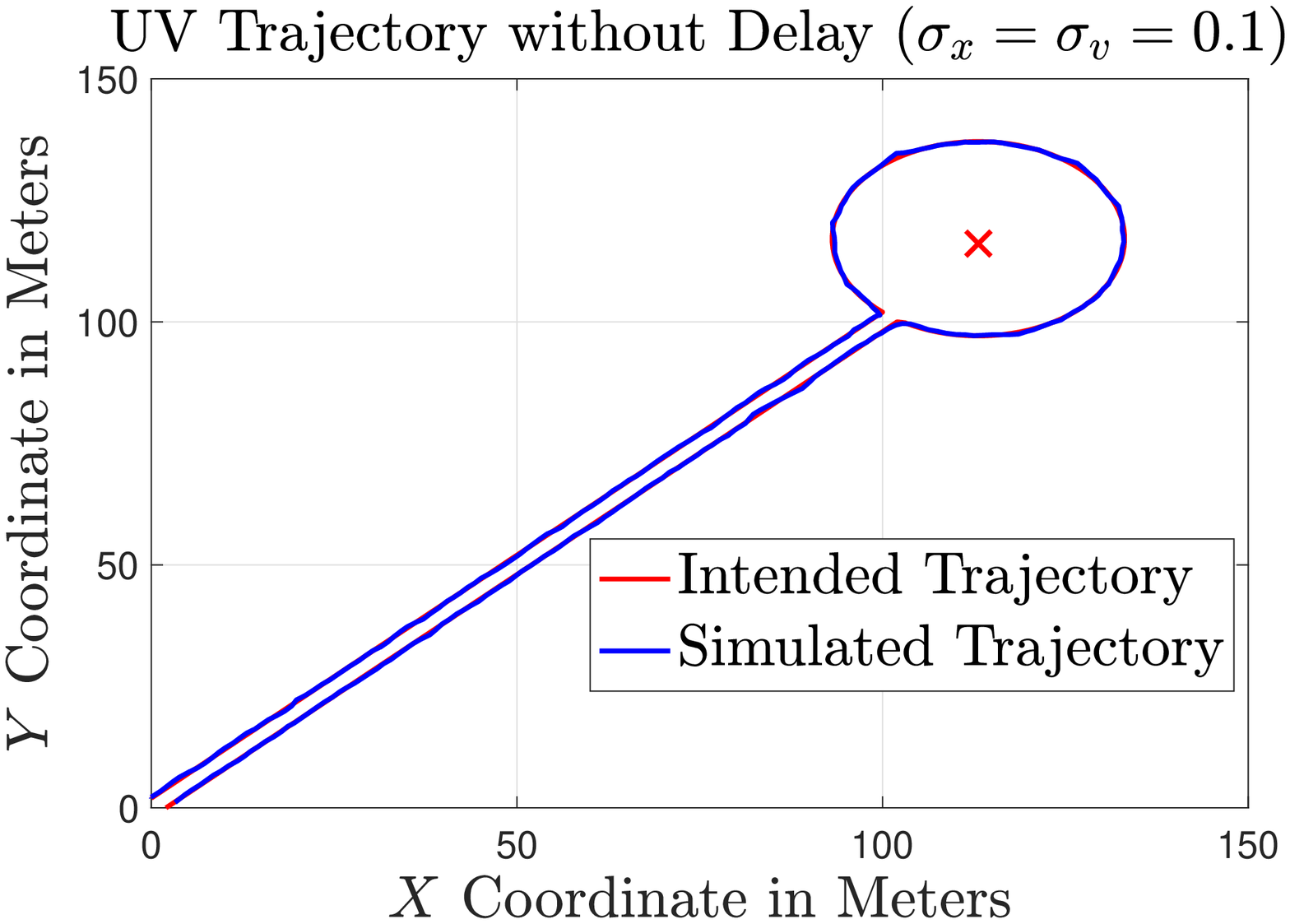}
\hspace{-0.5cm}
\includegraphics[scale=0.23]{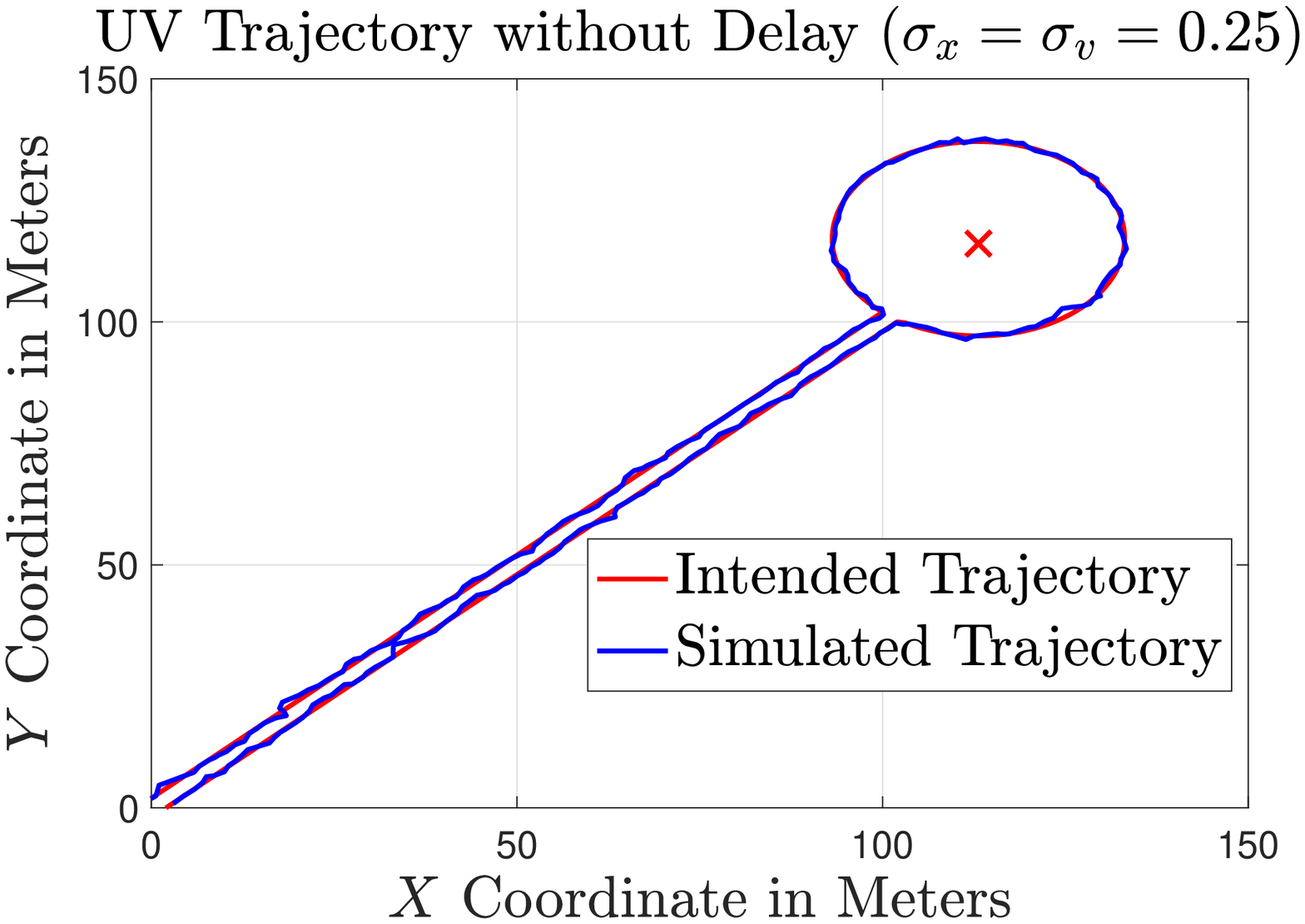}
\hspace{-0.5cm}
\includegraphics[scale=0.23]{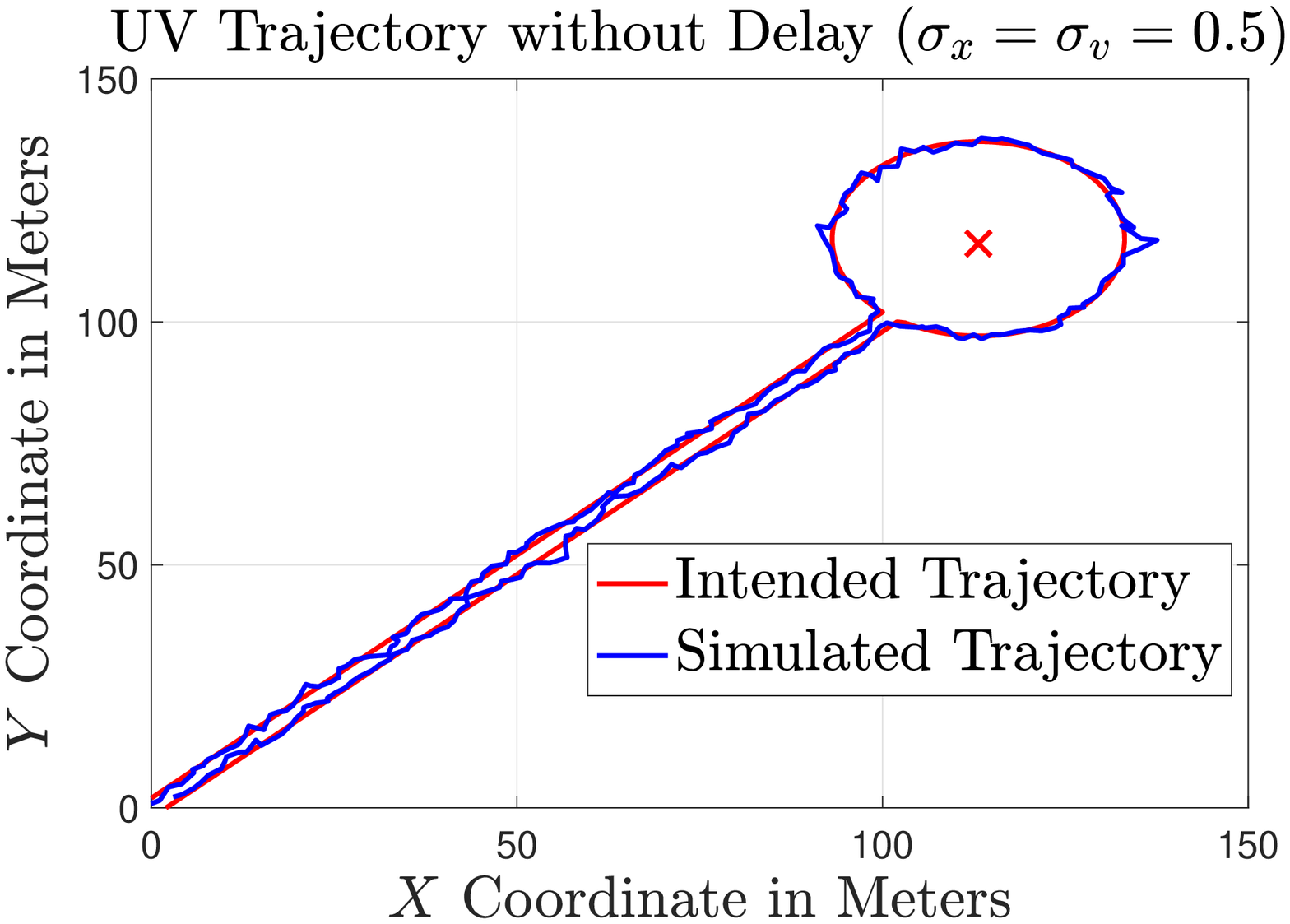}
\hspace{-0.5cm}
\includegraphics[scale=0.23]{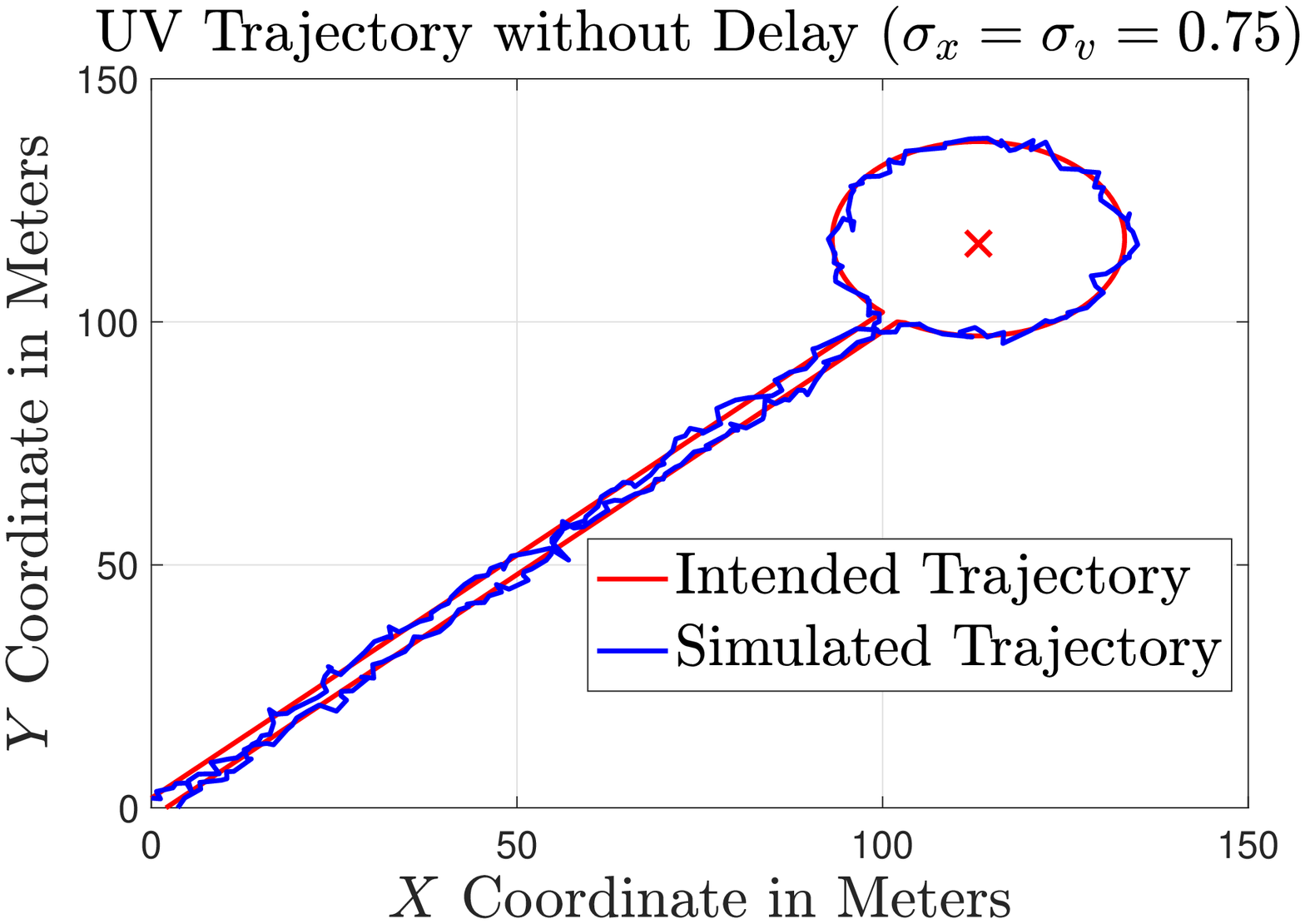}
\end{center}
\end{minipage}
\\
\begin{minipage}[t]{20cm}
\begin{center}
\hspace{-1.75cm}
\includegraphics[scale=0.23]{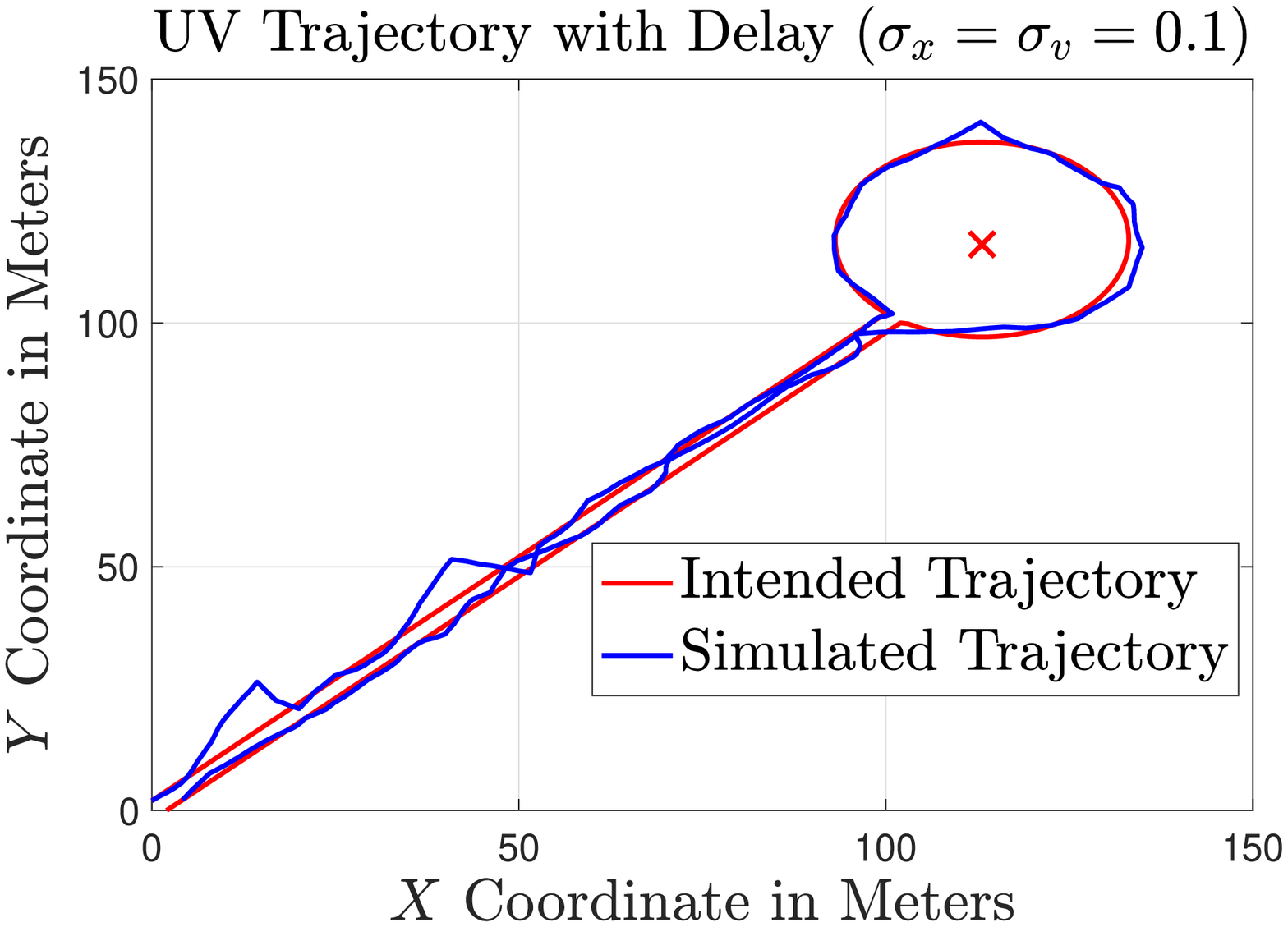}
\hspace{-0.5cm}
\includegraphics[scale=0.23]{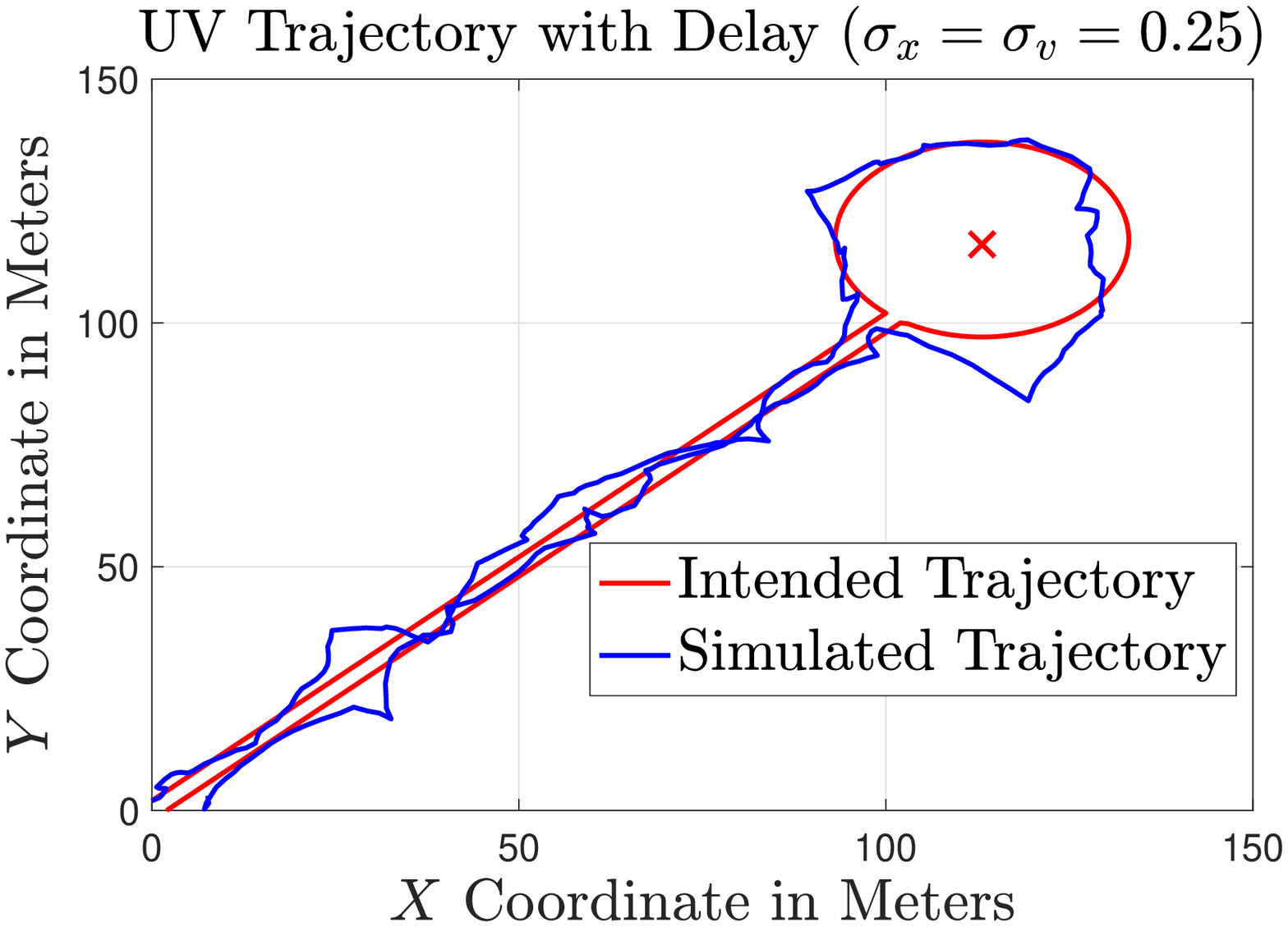}
\hspace{-0.5cm}
\includegraphics[scale=0.23]{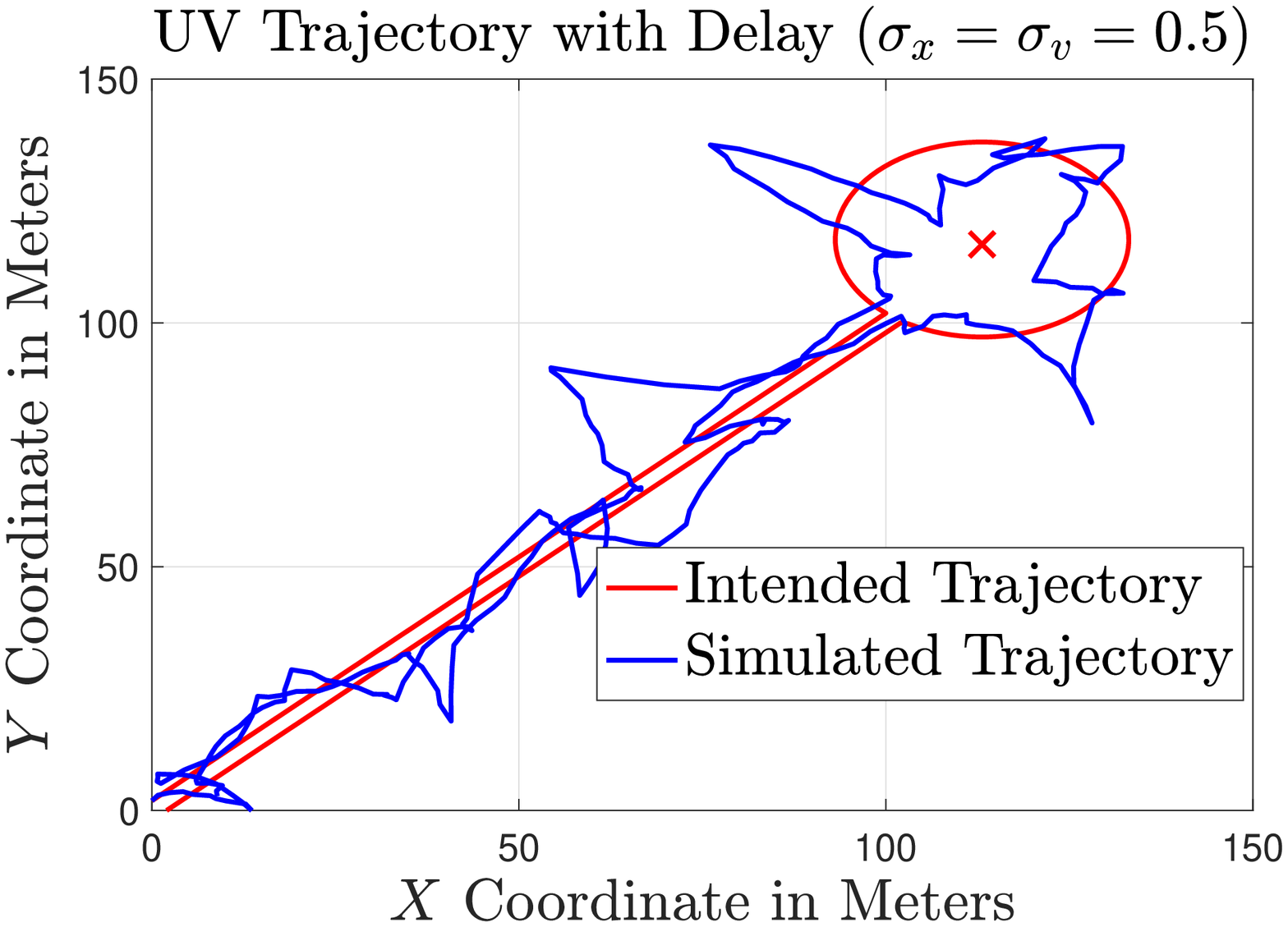}
\hspace{-0.5cm}
\includegraphics[scale=0.23]{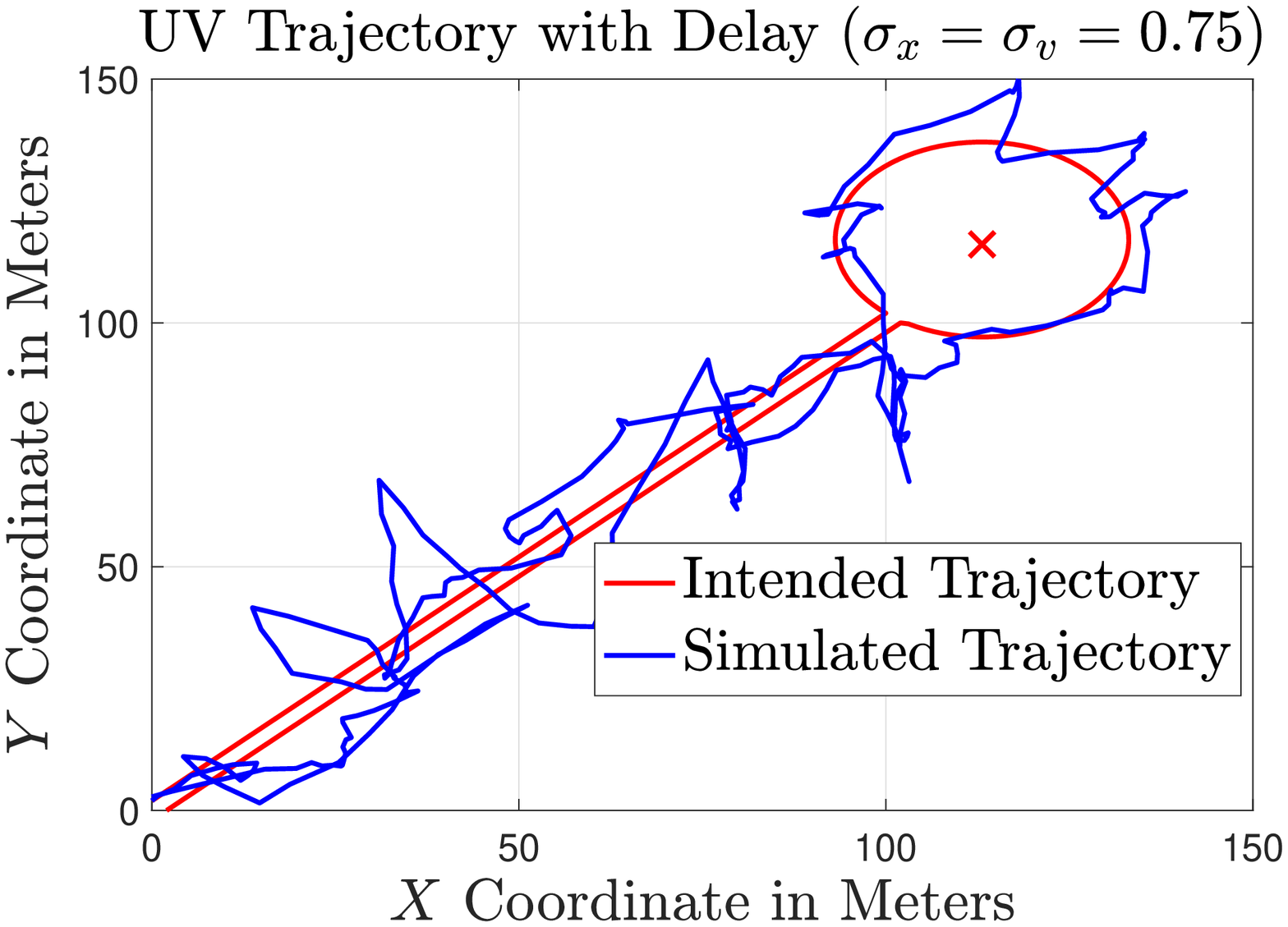}
\end{center}
\end{minipage}
\caption{Effect of environmental disturbances on the trajectory tracking performance of a UV fog controller. ($p = 0.75$, $\alpha = 0.1$ and $\rho = \frac{\sigma_x \sigma_v}{2}$ for all figures. The upper figures are for the case without delay, whereas the lower figures are for the case with delay $M = 3 \Delta t$.)}  \label{Fig: Figure 3}
\end{figure*}

Second, we investigate the effect of environmental disturbances on the trajectory tracking performance of a UV fog controller in Fig. \ref{Fig: Figure 3}.  We set $p$ to $0.75$ and $\alpha$ to $0.1$, and chose $\rho$ to have a correlation coefficient of $0.5$ in all cases.  In the upper figures, we again observe that our optimum fog controller is able to stabilize the UV around the predefined desired trajectory, albeit having more jitters around the way-points with harsher environmental conditions. In particular, we see that there is a constant fight between efforts from the fog controller to track the desired path and random disturbances due to environmental conditions such as wind to deviate from the desired path.  In all cases, the UV does never go uncontrolled, which is a positive indication for the perfectly matched fog controller from the perspective of preventing potential collisions with other vehicles or obstacles existing in close geographical distances.  In the lower figures in Fig. \ref{Fig: Figure 2}, on the other hand, we observe that the jitters due to environmental disturbances are more pronounced for the fog controller with delay.  The UV motion almost resembles a random walk around the desired trajectory, which is certainly not desirable for many mission critical applications.  This observation is mainly because of the accumulation of random disturbances until a delay-spread corrective action against trajectory deviations is taken by the fog controller.  These results signify the importance of dynamic provisioning of quality of control service over fog through a lever to adjust fog node reliability and delay, especially in cases when the jitter around a desired trajectory is detrimental for the mission executed by the UV such as surveillance monitoring of a geographical region, a remote first-aid operation and a remote swimmer rescue operation by means of drones.

\begin{figure*}[!t]
\begin{minipage}[t]{20cm}
\begin{center}
\hspace{-1.75cm}
\includegraphics[scale=0.23]{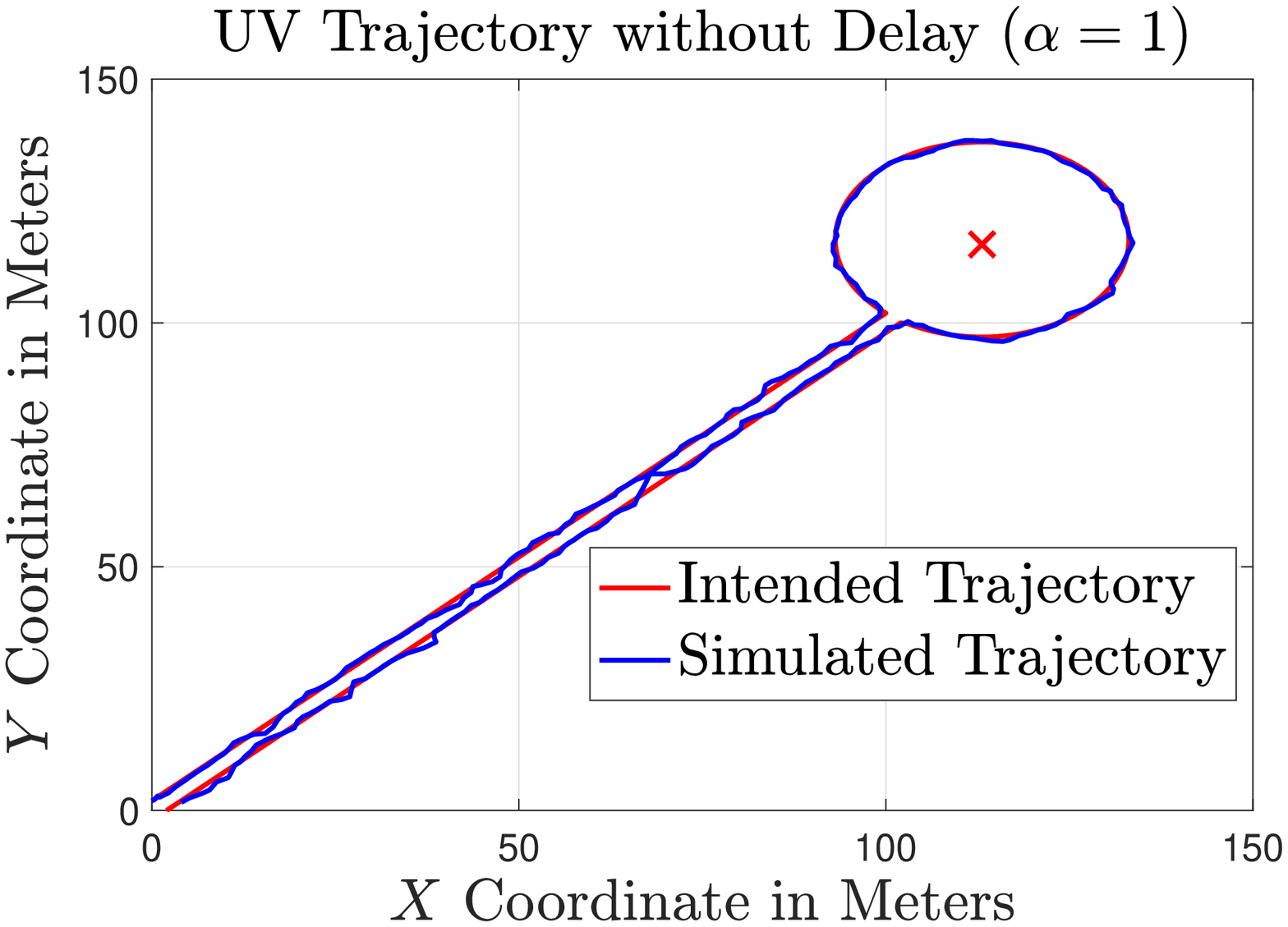}
\hspace{-0.5cm}
\includegraphics[scale=0.23]{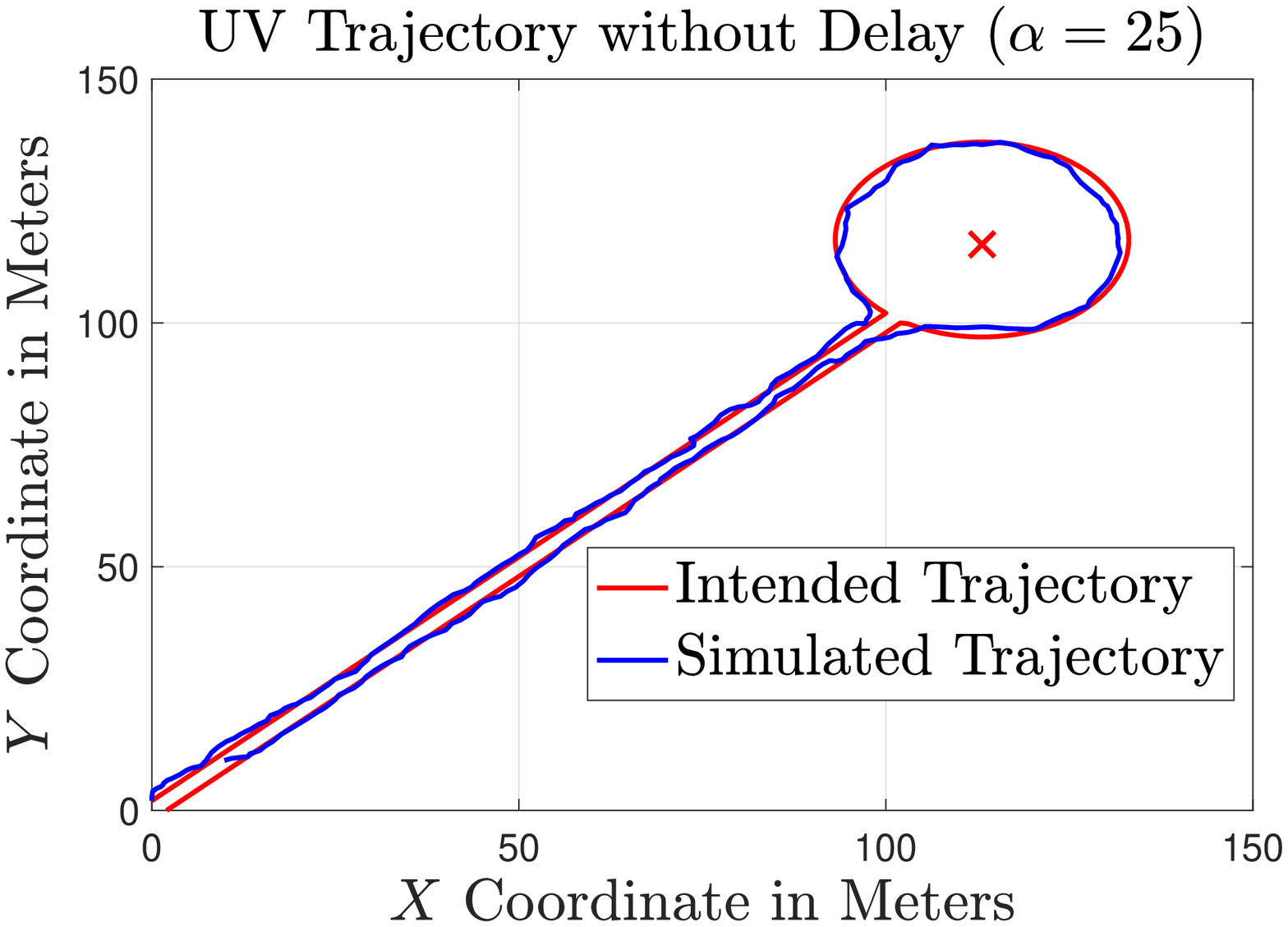}
\hspace{-0.5cm}
\includegraphics[scale=0.23]{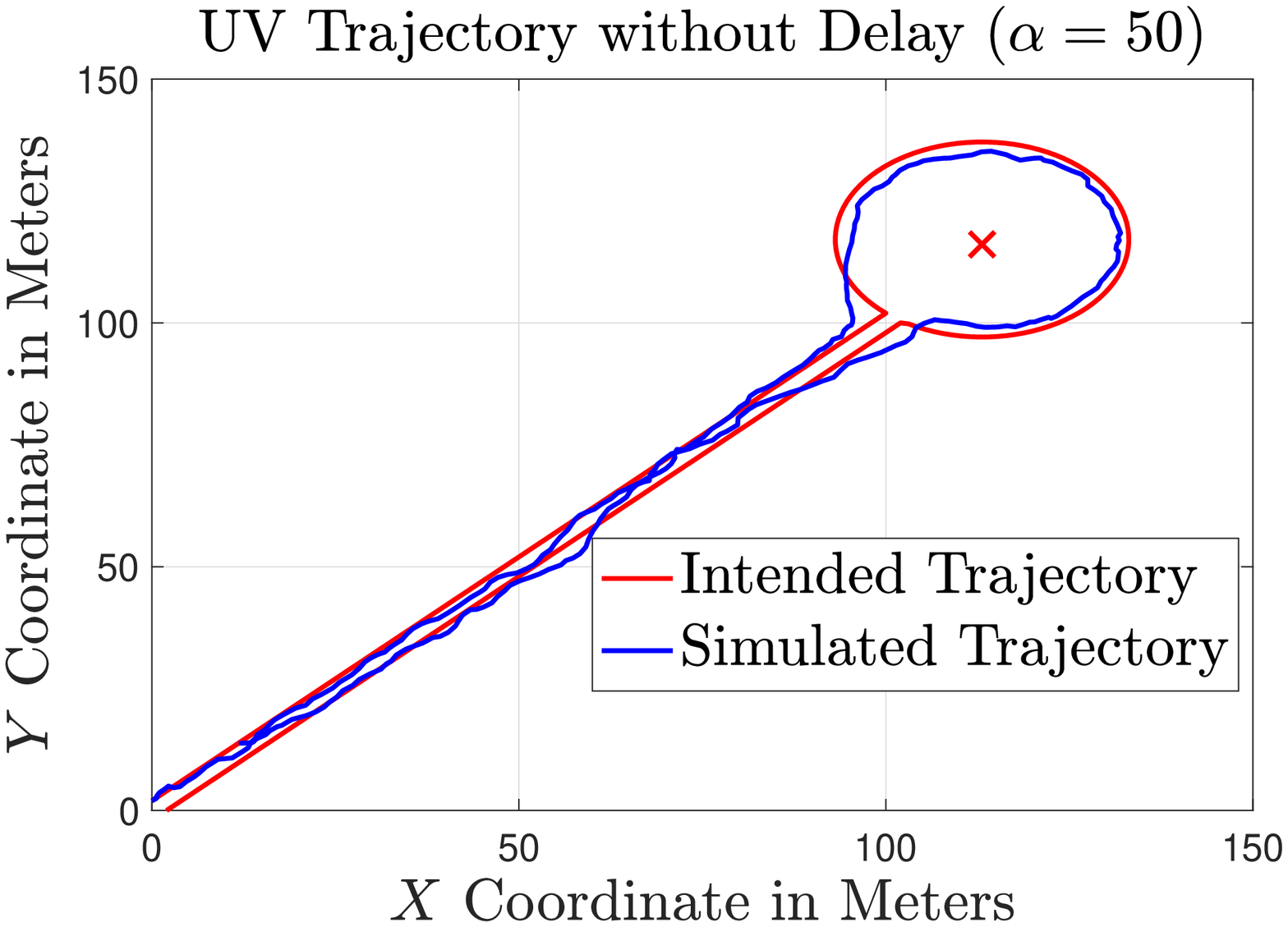}
\hspace{-0.5cm}
\includegraphics[scale=0.23]{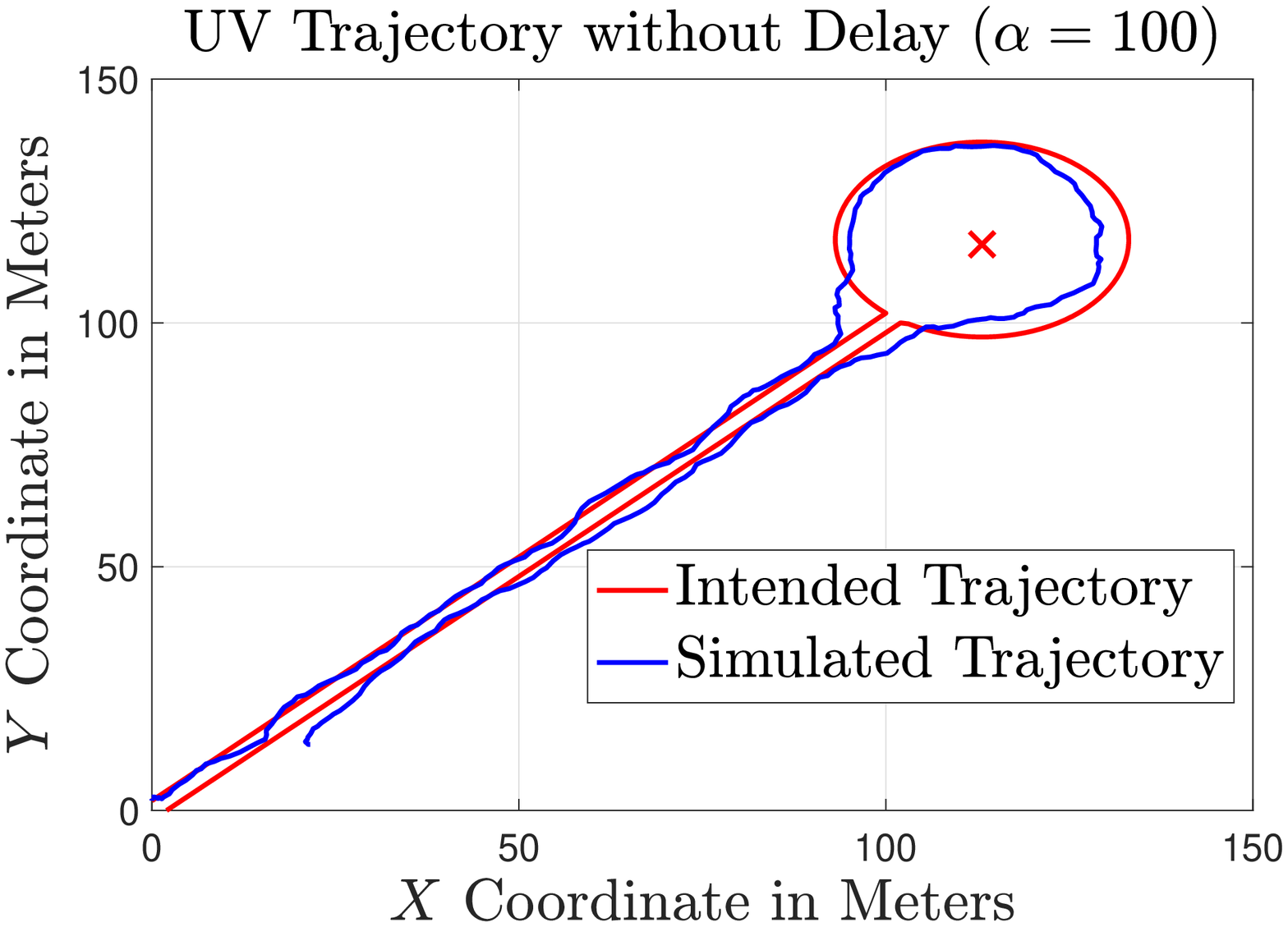}
\end{center}
\end{minipage}
\\
\begin{minipage}[t]{20cm}
\begin{center}
\hspace{-1.75cm}
\includegraphics[scale=0.23]{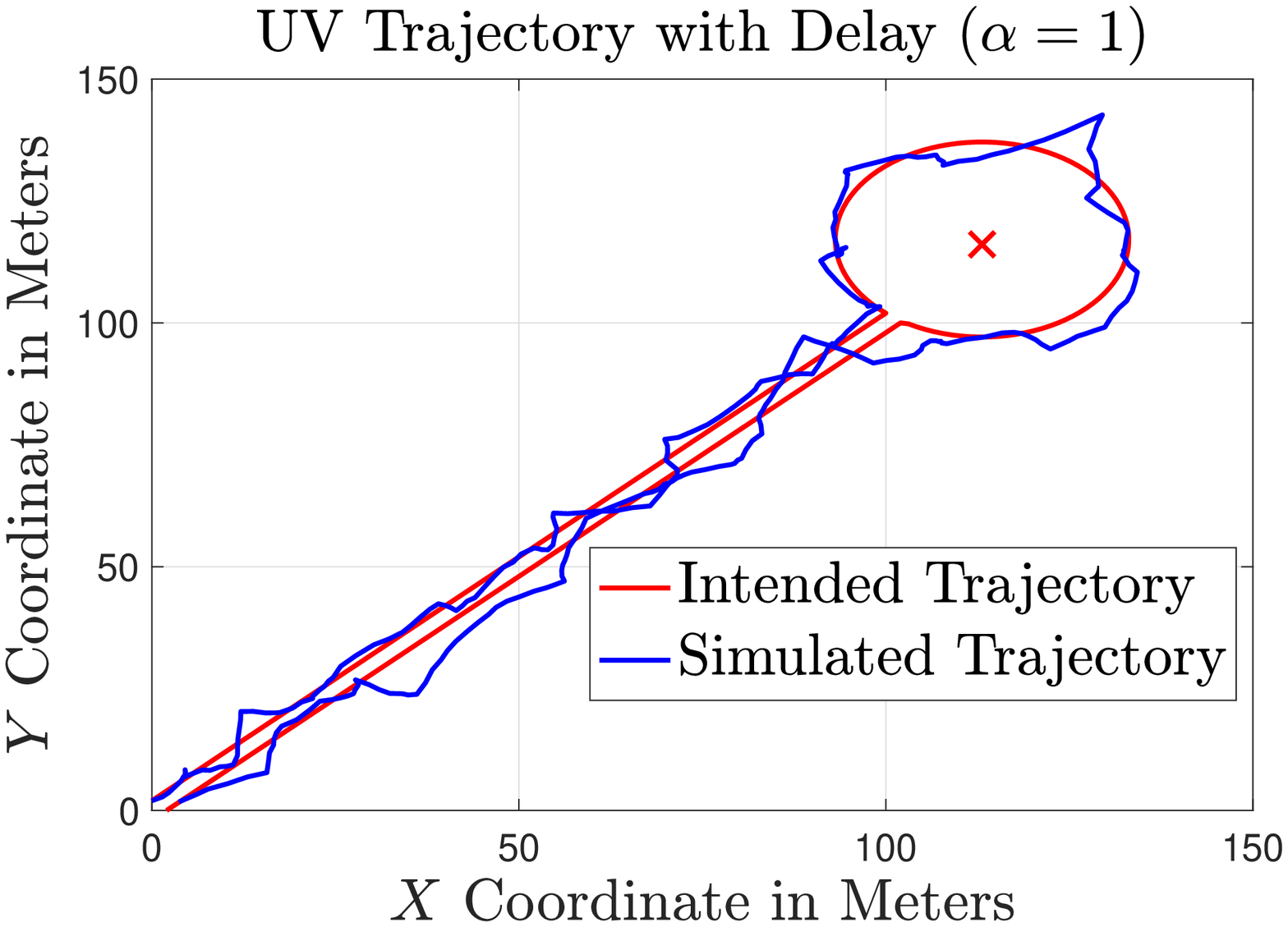}
\hspace{-0.5cm}
\includegraphics[scale=0.23]{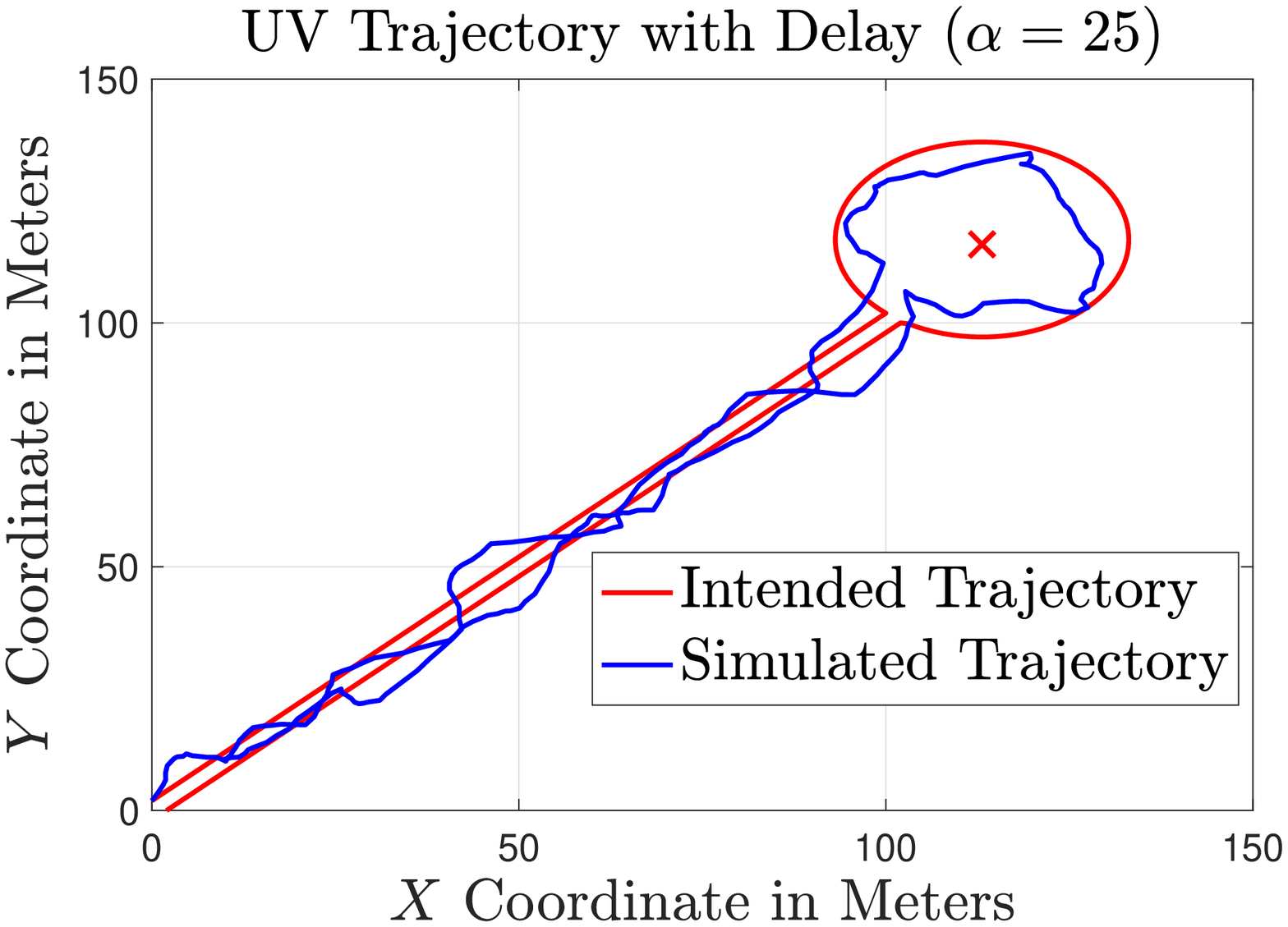}
\hspace{-0.5cm}
\includegraphics[scale=0.23]{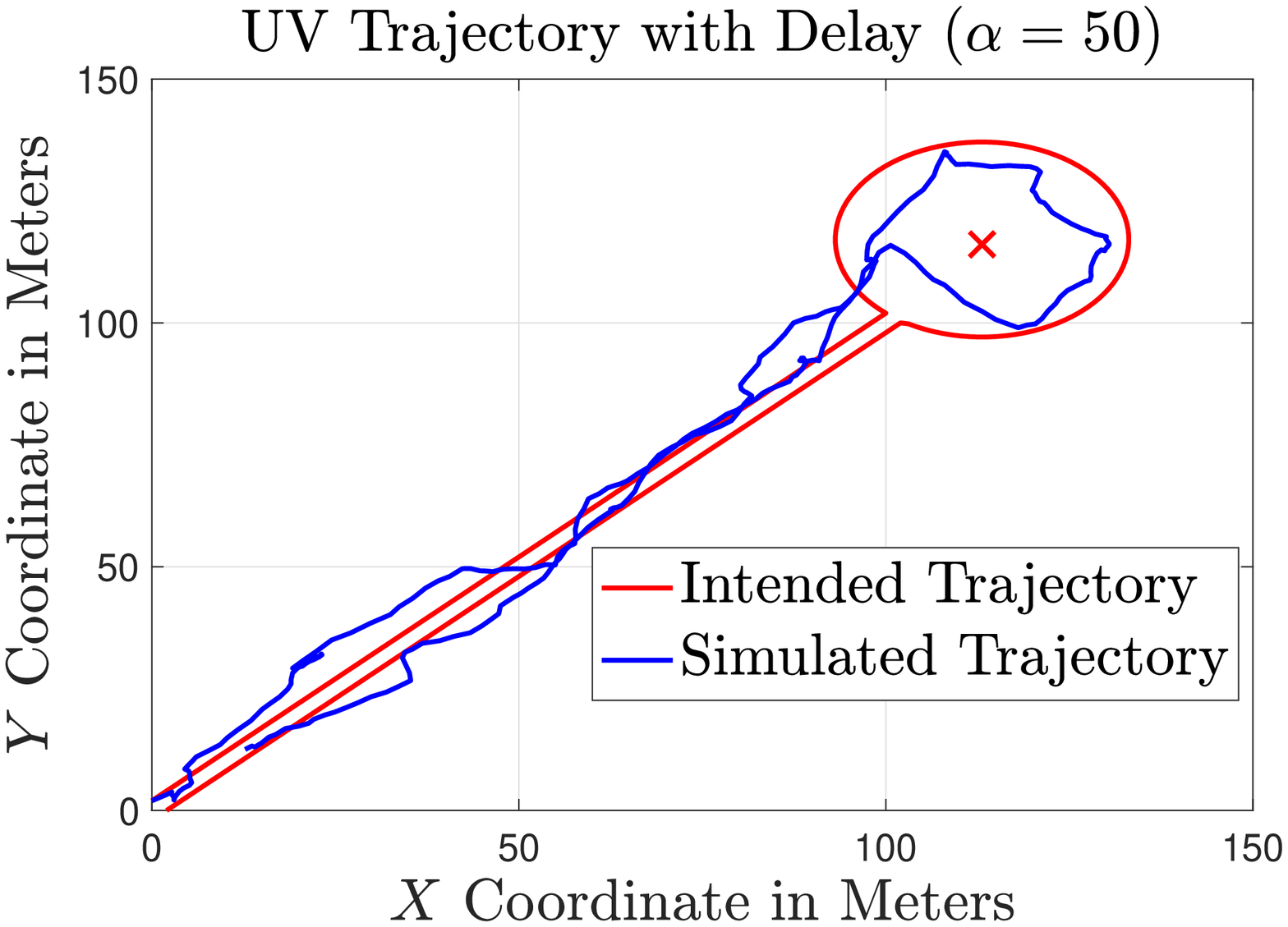}
\hspace{-0.5cm}
\includegraphics[scale=0.23]{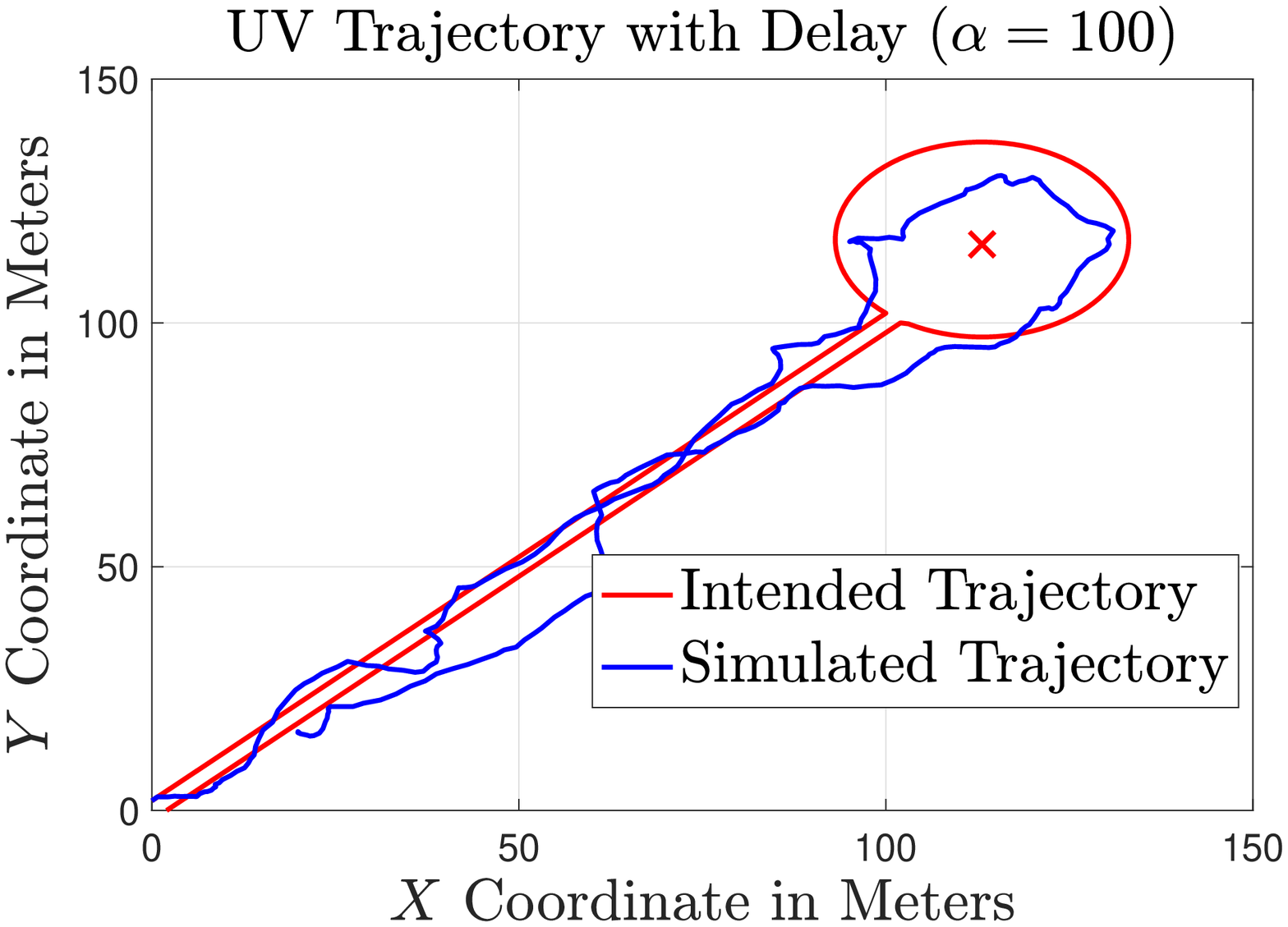}
\end{center}
\end{minipage}
\caption{Effect of the energy weight parameter $\alpha$ on the trajectory tracking performance of a UAV fog controller. ($p = 0.75$, $\sigma_x = 0.25$, $\sigma_v = 0.25$ and $\rho = \frac{\sigma_x \sigma_v}{2}$ for all figures. The upper figures are for the case without delay, whereas the lower figures are for the case with delay $M = 3 \Delta t$.)}  \label{Fig: Figure 4}
\end{figure*}

In Fig. \ref{Fig: Figure 4}, we study the effect of the system-level parameter $\alpha$ on the trajectory tracking performance of the UV fog controller.  In this figure, we set $p$, $\sigma_x$ and $\sigma_v$ to $0.75$, $0.25$ and $0.25$, respectively.  We chose $\rho$ to have a correlation coefficient of $0.5$ in all cases.  The parameter $\alpha$ helps us to adjust the weight associated with the total energy spent during the journey of the UV around the desired trajectory. By increasing the value of $\alpha$, the importance ranking of the energy spent rises with respect to the relative importance of how well the UV tracks its trajectory.  This change of perspective results in deteriorations in the trajectory tracking performance of the UV controlled over fog, as illustrated in Fig. \ref{Fig: Figure 4}.  In particular, the already poor tracking performance of the UV fog controller with delay becomes much worse as such the UV deviates significantly from the desired trajectory with increasing values of $\alpha$ (i.e., smaller distorted circles around the target and incomplete paths) when there is delay to communicate measurements and control signals between the virtual fog controller and the velocity control unit of the UV.  An important engineering implication of these results is the more evident importance of the delay when the energy is at stake to control a UV over the fog.

\section{Conclusions, Future Work and Discussions} \label{Section: Conclusions}
Fog computing presents two levers of reliability and latency to regulate the performance of virtual control services to enable/manage smarter IoT endpoints over a network. In this paper, we have introduced a framework to investigate the potential of fog computing for this end. Specifically, we have derived optimum control policies and the resulting min-cost performance for controlling stochastic IoT node processes by considering service reliability and communication/computation latency over a fog network. Our results reveal the way in which reliability and latency influence the quality of virtual control services over fog. In particular, it has been observed that latency is more critical than reliability in a fog computing environment since it determines both the frequency of corrective control actions and the timeliness of state measurements. These results have been illustrated for a drone trajectory tracking control problem.

This work offers an initial step to discover the utility of fog computing for virtual control applications, with several important future research directions remaining. Firstly, this paper does not consider how to provision control services for multi-tenant control applications running at the same fog endpoint. In such cases, a performance criterion must be jointly optimized over multiple clients by considering their service blockage probabilities and latencies. Secondly, an imperfect match between the virtual fog controller and the IoT node process introduces only an initial setup delay without impacting the frequency of corrective control to a large extent in some applications. In such cases, our analysis presented in Section \ref{Section: Control With Latency} still applies, but with a virtual estimator and controller running at all times after the initial setup delay. Over a large time-horizon, the negative effects of initial setup delay can be counteracted and the reliability may emerge more detrimental than latency in these cases. Finally, extensions of our results in this paper to non-linear IoT node processes are also of interest for control applications in which the linearity assumption in \eqref{Fig: System Model} does not hold or a reasonable linear approximation for the IoT node process cannot be obtained.





\appendices
\section{Proof of Theorem \ref{Thm: Min Cost Bounds1}} \label{Appendix: Bounds 1}
Consider two different systems, the first one with Markov transition probabilities $1-q$ and $q$, and the second one with $p$ and $q$. Consider the optimum control $\vec{\pi}^\star$ achieving $\mincost{1-q}{q}$ for the first system. We apply the same control to System $2$, although being sub-optimum, to obtain the upper bound. Let $\mingocost{1-q}{q}{k}$ and $\gocost{p}{q}{k}$ be the cost-to-go values for systems $1$ and $2$ under $\vec{\pi}^\star$, starting at $\vec{x}_k$ and $\tau_k$. It is easy to see that
\begin{eqnarray}
\gocostON{p}{q}{k} \leq \mingocostON{1-q}{q}{k}
\end{eqnarray}
and
\begin{eqnarray}
\gocostOFF{p}{q}{k} \leq \mingocostOFF{1-q}{q}{k}
\end{eqnarray}
for $k=N-1, N$. Assume the same holds for $k+1 \leq N-1$ as an inductive argument. Then, we have
\begin{eqnarray*}
\lefteqn{\gocostON{p}{q}{k} = \vec{x}_k^\top \vec{Q}_k \vec{x}_k + \paren{\vec{u}_k^\star}^\top \vec{R}_k \vec{u}_k^\star} \hspace{8.0cm} \\
\lefteqn{+ p \ESI{\gocostON{p}{q}{k+1} \big| \vec{u}_k^\star}{\paren{\vec{x}_k, 1}}} \hspace{6.0cm} \\
\lefteqn{+ (1-p) \ESI{\gocostOFF{p}{q}{k+1} \big| \vec{u}_k^\star}{\paren{\vec{x}_k, 1}},} \hspace{6.0cm}
\end{eqnarray*}
where $\vec{u}_k^\star$ is the optimum control applied to System $1$ at time $k$ after observing $\vec{x}_k$. Using the inductive argument and observing that $\mingocostON{1-q}{q}{k} \leq \mingocostOFF{1-q}{q}{k}$, we have
\begin{eqnarray*}
\lefteqn{\gocostON{p}{q}{k} \leq \vec{x}_k^\top \vec{Q}_k \vec{x}_k + \paren{\vec{u}_k^\star}^\top \vec{R}_k \vec{u}_k^\star} \hspace{8.0cm} \\
\lefteqn{+ q \ESI{\mingocostOFF{1-q}{q}{k+1} \big| \vec{u}_k^\star}{\paren{\vec{x}_k, 1}}} \hspace{6.0cm} \\
\lefteqn{+ (1-q) \ESI{\mingocostON{1-q}{q}{k+1} \big| \vec{u}_k^\star}{\paren{\vec{x}_k, 1}}} \hspace{6.0cm} \\
\lefteqn{= \mingocostON{1-q}{q}{k}.} \hspace{6.0cm}
\end{eqnarray*}
Repeating the same steps for $\gocostOFF{p}{q}{k}$ shows that $ \mingocost{p}{q}{k} \leq \mingocost{1-q}{q}{k}$ since $\mingocost{p}{q}{k} \leq \gocost{p}{q}{k}$, which proves the upper bound. Similar arguments apply for the lower bound, too.

\balance
\bibliographystyle{ieeetr}
\bibliography{ResearchStatement2016}

\end{document}